\documentstyle[12pt]{amsart}

\newcommand{\A}{\cal{A}}
\newcommand{\As}{\cal As}
\newcommand{\B}{\cal{B}}

\newcommand{\Bh}{{\B'}}

\newcommand{\cc}{\cal{C}}
\newcommand{\cch}{{\cc'}}
\newcommand{\C}[2]{C_{#1}^{#2}}
\newcommand{\Comm}{\cal Comm}
\newcommand{\CP}[1]{\Bbb{C}\Bbb{P}^{#1}}
\newcommand{\D}{\cal{D}}
\newcommand{\e}[1]{{\Theta}_{#1}}
\newcommand{\End}[1]{\cal{E}@!n@!d\,_{#1}\,}

\newcommand{\G}{\cal{G}}
\newcommand{\GG}{\widetilde{\G}}
\newcommand{\Hom}{\operatorname{Hom}}
\newcommand{\id}{\operatorname{id}}
\newcommand{\ie}{{\sl i.e.}}
\newcommand{\into}{\hookrightarrow}
\newcommand{\identity}{{\bf e}}
\newcommand{\I}{\bf{I}}
\newcommand{\In}{\operatorname{In}}
\newcommand{\K}{\cal{K}}
\newcommand{\Lie}{\cal Lie}
\newcommand{\nc}{{\Bbb{C}}}
\newcommand{\nr}{{\Bbb{R}}}
\newcommand{\oo}{\circ}
\newcommand{\OO}{\cal{O}}
\newcommand{\op}{{\operatorname{op}}}
\newcommand{\p}{\cal{P}}
\newcommand{\pp}{\widetilde{\cal{P}}}
\newcommand{\PSL}{\operatorname{PSL}(2,\nr)}
\newcommand{\PSLc}{\operatorname{PSL}(2,\nc)}
\newcommand{\RP}[1]{\Bbb{R}\Bbb{P}^{#1}}
\newcommand{\Sigmat}{\widetilde{\Sigma}}
\newcommand{\vacuum}{\bf{\Lambda}}

\newtheorem{th}{Theorem}
\newtheorem{lm}[th]{Lemma}
\newtheorem{prop}[th]{Proposition}

\theoremstyle{definition}

\newtheorem{df}{Definition}
\newtheorem{ex}{Example}

\theoremstyle{remark}

\newtheorem{rem}{Remark}

\newenvironment{ack}{\small \trivlist \item[\hskip \labelsep{\it
Acknowledgments}.]}{\endtrivlist}

\begin{document}

\title{The cohomology of algebras over moduli spaces}

\author
{Takashi Kimura}
\address
{Dept. of Mathematics, University of North Carolina, Chapel Hill, NC
27599-3250, USA}
\email{kimura@@math.unc.edu}
\thanks{Research of the first author was supported in part by an NSF
Postdoctoral Research Fellowship}

\author
{Alexander A. Voronov}
\address
{Department of Mathematics, University of Pennsylvania, Philadelphia, PA
19104-6395, USA}
\email{voronov@@math.upenn.edu}
\thanks{Research of the second author was supported in part by an NSF grant}

\date{October 28, 1994}

\maketitle

The purpose of this paper is to introduce the cohomology of various
algebras over an operad of moduli spaces including the cohomology of
conformal field theories (CFT's) and vertex operator algebras
(VOA's). This cohomology theory produces a number of invariants of
CFT's and VOA's, one of which is the space of their infinitesimal
deformations.

The paper is inspired by the ideas of Drinfeld \cite{dr}, Kontsevich
\cite{kon:sympl} and Ginzburg and Kapranov \cite{gk} on Koszul duality for
operads.  An operad is a gadget which parameterizes algebraic operations on a
vector space. They were originally invented \cite{m} in order to study the
homotopy type of iterated loop spaces. Recently, operads have turned out to be
an effective tool in describing various algebraic  structures that arise in
mathematical physics in terms of the geometry of  moduli spaces particularly in
conformal field theory (see \cite{g}, \cite{h}, \cite{ksv}, \cite{lz},
\cite{wz}, \cite{z}) and topological gravity (see \cite{ksv2}, \cite{konm}). In
fact, a (tree level $c=0$) conformal field theory is nothing more than a
representation of the operad, $\p$, consisting of moduli spaces of
configurations of holomorphically embedded unit disks in the Riemann sphere.
Alternately, a conformal field theory is said to be a $\p$-algebra or an
algebra over $\p$.

In the first part of this paper, we use the idea of Ginzburg-Kapranov
\cite{gk} of homology of an algebra over a quadratic operad to define
the {\sl co}homology of an algebra over a quadratic operad with
values in an arbitrary module. We prove that $\p$ is a quadratic
operad and construct its Koszul dual operad. We then construct the
cohomology theory associated to a conformal field theory. We
demonstrate that the second cohomology group, as it should,
parameterizes deformations of conformal field theories. Our approach
to deformations is morally the same as the one developed by Dijkgraaf
and E. and H. Verlinde \cite{dijk,dvv} and Ranganathan, Sonoda and Zwiebach
\cite{sz} using (1,1)-fields, except that we fix the action of the Virasoro
algebra. It would be very interesting to find an explicit connection
between the two approaches. By analogy with the case of associative
algebras, one expects the third cohomology group to contain
obstructions to extending an infinitesimal deformation to a formal
neighborhood. Another plausible interpretation of higher dimensional
cohomology is that they should parameterize infinitesimal deformations
inside a larger category of ``$A_\infty$-conformal field theories'' where one
allows multilinear vertex operators which are not compositions of bilinear
ones.

The second part of this paper is structured in the opposite way. We
construct the cohomology theory associated to algebras over the little
intervals operad, $\B$ -- the space of configurations of intervals
embedded in the interval via translations and dilations -- by studying
deformations of $\B$-algebras in analogy with the manner in which
Hochschild cohomology arises from deformations of associative
algebras. We do so by realizing associative algebras as one
dimensional topological field theories. The little intervals operad is
an important suboperad of the real analytic analog of $\p$ which can
be regarded as a suboperad of $\p$. Therefore, a conformal field
theory is a $\B$-algebra. This leads us to a complex which, in the
case where the conformal field theory is a so-called vertex operator
algebra, can be written explicitly. This complex closely resembles
Hochschild cohomology and looks like what one would obtain from
formally deforming the operator product in a
vertex operator algebra.

Throughout the paper, we restrict our attention to tree level
theories, those which correspond to Riemann surfaces of genus zero.
There should be a cyclic version of the cohomology theory considered
here, as it happens for the homology of algebras over cyclic operads,
see \cite{gek}.

A computation of the cohomology of a conformal field theory would
provide valuable information about the structure of the moduli space
of conformal field theories which plays an important role in related
phenomena such as mirror symmetry, cf.\ Kontsevich and Manin
\cite{konm}. We expect techniques from the theory of vertex operator
algebras to be useful in performing this computation.

\subsection{Algebras and operads}

Assume for simplicity that all vector spaces are over $\nc$. Perhaps
the simplest example of an operad is the {\sl endomorphism operad} of
a vector space $V$, denoted by $\End{V} = \{\,\End{V}(n)\,\}_{n\geq 1}$,
which is a collection of spaces $\End{V}(n) = \Hom(V^{\otimes n},V)$, each
with the natural action of the permutation group, $S_n$, which acts  by
permuting the factors of the tensor product, and which have natural
compositions between them, namely, the composition, $f \oo_i f'$, of two
elements $f$ in $\OO(n)$ and $f'$ in $\OO(n')$ is obtained by
\begin{multline*}
(f\oo_i f')(v_1\otimes \cdots\otimes v_{n+n'-1}) \\
= f(v_1,\otimes\cdots\otimes
v_{i-1}\otimes f'(v_i\otimes\cdots v_{i+n'-1})\otimes\cdots\otimes v_{n+n'-1})
\end{multline*}
for all $i=1,\ldots, n$, and the permutation groups act equivariantly
with respect to the compositions. Finally, there is an element $\I$ in
$\End{V}(1)$, the identity map $\I:V\,\to V$, which is a unit with
respect to the composition maps. This structure can be formalised in the
following way.

An {\sl operad  $\OO = \{\,\OO(n)\,\}_{n\geq 1}$ with unit} is a collection
of objects (topological spaces, vector spaces, etc. -- elements of any
symmetric monoidal category) such that each $\OO(n)$ has an action of $S_n$,
the permutation group on $n$ elements, and a collection of operations for
$n\geq 1$ and $1\leq i\leq n$, $\OO(n)\times\OO(n')\,\to\,\OO(n+n'-1)$ given
by $(f,f')\,\mapsto\,f\oo_i f'$ such that

\begin{enumerate}
\item if $f\in\OO(n),$ $f'\in\OO(n')$, and $f''\in\OO(n'')$ where $1\leq i <
j \leq n$ then
\begin{equation*}
(f\oo_i f')\oo_{j+n'-1} f'' = (f\oo_j f'')\oo_i f'
\end{equation*}
\item if $f\in\OO(n),$ $f'\in\OO(n')$, and $f''\in\OO(n'')$ where $n,n'\geq
1$ and $i = 1, \ldots, n$ and $j = 1, \ldots, n'$ then
\begin{equation*}
(f\oo_i f')\oo_{i+j-1} f'' = f\oo_i (f'\oo_j f''),
\end{equation*}
\item the composition maps are equivariant under the action of the
permutation groups,
\item there exists an element $\I$ in $\OO(1)$ called the {\sl
unit} such that for all $f$ in $\OO(n)$ and $i=1,\ldots,n$,
\begin{equation*}
\I\oo_1 f = f = f\oo_i \I
\end{equation*}
\end{enumerate}

This definition of an operad can be generalized by including $\OO(0)$ or by
relaxing the condition that the unit element exist but at the cost of having to
introduce additional axioms. We shall see that many of the constructions
which follow are quite at ease in this more general setting.

Given an operad, there is a notion of a representation of this operad on a
vector space. Let $\OO$ be an operad, a vector space $V$ is said to be an
{\sl $\OO$-algebra} if there is a morphism of operads $\mu:\OO\,\to\,
\End{V}$. That is to say, there exists a map $\mu:\OO(n)\,\to\,\End{V}(n)$,
one for each $n\geq 1$, such that $\mu$ preserves all of the structures.

There is a natural notion of a {\sl $V$-module, $M$, where $V$ is an
$\OO$-algebra.} The composition maps of a $V$-module are defined by taking the
axioms satisfied by the composition maps of an $\OO$-algebra and replacing
the $V$ associated to one input and the output with $M$, and demanding that
the axioms of an algebra hold (see \cite{gk}) except that one does not allow
compositions between elements in $M$. In other words, the axioms of a
$V$-module $M$ are those which make the algebra $V$ itself naturally into a
$V$-module.

\section{Conformal Field Theory and Vertex Operator Algebras}

A ($c=0$ tree level) conformal field theory is an algebra over an operad of
moduli spaces of configurations of holomorphically embedded disks in a Riemann
sphere.

Let $D$ be the (closed) unit disk in the complex plane. Let $\p(n)$ be the
moduli space of configurations of $(n+1)$-distinct, ordered biholomorphic
embeddings of $D$ into the Riemann sphere, $\CP{1} = \nc\cup\{\infty\}$ (in
a standard coordinate $z$) whose images may overlap at the boundaries of the
disks where any two such configurations are identified if they are related
by a automorphism of the Riemann sphere, \ie\ a complex projective
transformation, where $\PSLc$ acts upon $\CP{1}$ by
\begin{equation*}
z  \mapsto \, \frac{az+b}{cz+d},\ \forall\, a,b,c,d\in\nc\
\text{satisfying}\  \det\left(\matrix a & b\\ c & d \endmatrix\right) = 1.
\end{equation*}
$\p=\{\,\p(n)\,\}_{n\geq 1}$ forms an operad of complex manifolds where the
permutation group acts by permuting the first $n$ embeddings and the
composition maps $\p(n)\times\p(n')\,\to\,\p(n+n'-1)$ taking
$(\Sigma,\Sigma') \,\mapsto\,\Sigma\oo_i\Sigma'$ is defined by cutting out
the $(n+1)$st disk of $\Sigma'$, the $i$th disk of $\Sigma$ and then sewing
across the boundary by $x \mapsto 1/x$, $x$ being a standard complex
coordinate on $D$.

A {\sl (tree level $c=0$) conformal field theory (CFT)}, $V$, is a
$\p$-algebra,  \ie\  $V$ is a topological vector space with a smooth morphism
of operads $m:\p\,\to\,\End{V}.$ An important class of CFT's are {\sl
holomorphic conformal field theories} which are CFT's such that the morphism,
$m$, is holomorphic. Tree level means that only genus zero Riemann
surfaces appear in $\p$. A general tree level conformal field theory is a
collection of smooth maps $m:\p\to\,\End{V}$ which are equivariant under the
action of the permutation group and such that compositions of elements in
$\p$ are mapped into compositions of their images in $\End{V}$ up to a
projective factor. In particular, $V$ is a projective $\p(1)$ module and, as
a consequence, is a projective representation of the Virasoro algebra with
central charge $c$. Thus, if $m$ is a morphism of operads then
$c=0$. However, in the general case, $V$ can be regarded as an honest algebra
over a larger operad involving determinant line bundles which fibers over
$\p$ (see \cite{s}). For simplicity, we shall restrict to $c=0$ CFT's
though it is a simple matter to generalize what follows to the more general
case.

There is another collection of moduli spaces closely related to $\p$. Let
$\pp(n)$ be the moduli space of configurations of $(n+1)$-distinct ordered
holomorphic coordinates (no two of which have coinciding centers) on $\CP{1}$
where any two such configurations are identified if they are related by a
complex projective transformation. The collection, $\pp :=
\{\,\pp(n)\,\}_{n\geq 1}$, satisfies all the axioms of an operad except that
one cannot always compose elements in $\pp$. An object which satisfies
the axioms of an operad whenever compositions are defined is called a {\sl
partial operad}, see Huang-Lepowsky \cite{hl}. The partial operad $\pp$
contains $\p$ as a suboperad since holomorphically embedded unit disks may be
regarded as holomorphic coordinates. Since both $\p$ and $\pp$ are operads of
complex manifolds, the composition maps of $\pp$ are uniquely determined by
analytic continuation of the composition maps of $\p$. The composition of an
element $S$ in $\pp(n)$ and $S'$ in $\pp(n')$, $S\oo_i S'$, can be obtained
by identifying their boundaries. Suppose there exists a number $r > 0$ such
the disk of radius $r$ about the center of the $(n'+1)$st coordinate and the
disk of radius $1/r$ about the center of the $i$th coordinate do not contain
the centers of any other coordinates. In that case, $S\oo_i S'$ is obtained
but cutting out these disks and sewing. If such an $r$ does not exist then
$S\oo_i S'$ is undefined.

It is natural to define $V$ to be an algebra over $\pp$ if there is a
morphism of partial operads $\pp\,\to\,\End{V}$. We will call holomorphic,
$\pp$-algebras $V$ {\it ($c=0$) vertex operator algebras}\ ({\it VOA})
although as shown by Huang and Lepowsky \cite{hl}, an actual VOA has a
somewhat subtler structure. Because $\p$ is a suboperad of $\pp$, VOA's are
themselves holomorphic conformal field theories and provide an important
class of examples of such theories.

\subsection{Moduli spaces $\p (n)$ as a quadratic operad}

The operad $\p(n)$, $n \ge 1$, of moduli spaces of Riemann spheres
with $n+1$ holomorphically embedded disks forms a quadratic operad in
the sense of Ginzburg-Kapranov \cite{gk}. This means that the operad
$\p(n)$ is the quotient $Q(K,E,R)$ of the free $K= \p(1)$-operad
$F(E)(n)$ by an ideal generated by defining relations $R$ which are all
quadratic, i.e., lie in $F(E)(3)$. Recall that the free operad is given by
$F(E)(n) = \bigoplus_{\text{binary $n$-trees $T$}} E(T)$ where $E(T) =
\bigotimes_{v \in T} E(\operatorname{In} v)$, $\operatorname{In} v$ denoting
the set of incoming edges of a vertex $v$, generated by $E= \p(2)$.

Let us describe generators and relations of $\p(n)$ in more detail: we
will need this in order to find the Koszul dual operad $\p^!(n)$ in the
next section. From now on, we will work in the category of vector
spaces. In particular, we consider the vector space generated by
$\p(n)$ instead of $\p(n)$ itself. The ground
algebra $K$, the space $E$ of generators and the space $R$ of
relations will be, thus, vector spaces, whose generators we are going
to describe.

The {\it ground algebra} $K=\p(1)$ is the ``semigroup'' algebra of the
distinguished Virasoro semigroup: as a vector space, it is generated
by cylinders, isomorphism classes of Riemann spheres with two
nonoverlapping (counted with boundary) holomorphically embedded
disks. Denote the center of the first disk by 0 and the center of the
second by $\infty$. The operad composition endows $K$ with the
structure of a semigroup: gluing the disk around $\infty$ on one
cylinder to the disk around 0 on the second one, along the
parametrizations of the disks.

The {\it space $E$ of generators} is generated by pairs of pants,
which are three-holed Riemann spheres. The symmetric group $S_2$ acts
on $E$ by interchanging the legs and $K$ acts on $E$ from the left by
gluing a cylinder along the circle around $\infty$ to the waist of the
pants and $K^2$ acts on $E$ from the right by gluing cylinders at
their disks centered at 0 to either of the legs.

Elements in the free operad generated by $E$, $F(E)$, have a nice
geometric realization. Elements in $F(E)(n)$ can be regarded as
elements in $\p(n)$, finite linear combinations of spheres with $n+1$
holes, together with a set of
homotopy classes of curves which cuts each sphere into pairs of
pants. Such a decomposition gives rise to a binary tree with a pair of
pants associated to each vertex such that the root of each vertex is
associated to the $3$rd hole (``the waist'') on the pants and the
other two (``the legs'') are associated to the two incoming edges.

The {\it space $R$ of relations} is the subtlest. First, we need to
describe the space $F(E)(3)$, where they live. As a vector space, it
is generated by the following elements. Each element is an equivalence
class of a triple $(T, p_1, p_2)$, where $T$ is a binary 3-tree, one
of the following three:
\begingroup
\input{psfig}
\begin{figure}[h]
\centerline{\psfig{figure=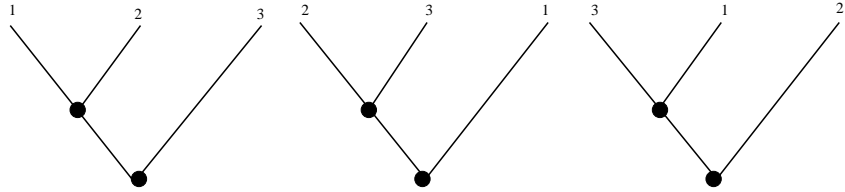,height=1in}}
\caption{Trees Indexing $F(E)(3)$}
\end{figure}
\endgroup

Here $p_1$ and $p_2$ are two pairs of pants, associated to the two
vertices of the tree $T$: $p_1$ to the lower vertex, $p_2$ to the
upper one. One can think of the triple $(T, p_1, p_2)$ as a Riemann
sphere with four holomorphic disks along with a curve cutting it into
two pairs of pants:

\begingroup
\input{psfig}
\begin{figure}[h]
\centerline{\psfig{figure=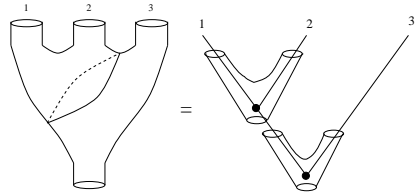,height=1in}}
\caption{An Element of F(E)(3) and Its Associated Tree}
\label{onetree}
\end{figure}
\endgroup

Two triples are considered equivalent if they differ by the following diagonal
action of $K$:
\[
(T, p_1 \circ_1 k, p_2) \sim (T, p_1, k \circ p_2) .
\]
Geometrically, this action slides the cut on the sphere with four
holes along the tube that it is wrapped around. Thus, the space
$F(E)(3)$ may be described as generated by isomorphism classes of
four-holed spheres with homotopy classes of cuts into two pairs of
pants.

The space $R$ of relations may be then described as follows:
\[
R = \bigoplus_{\left\{\parbox{.7in}{\scriptsize classes of 4-holed
spheres}\right\}}
\left\{\, \sum_{\gamma} a_\gamma e_\gamma \; \left| \;
\parbox{2in}{$a_\gamma = 0$  for all but a finite number of $\gamma$,
$\sum_\gamma a_\gamma =0$}\right. \right\},
\]
where the sum over $\gamma$ is the sum over homotopy classes of curves of
the 4-holed sphere which cut it into two pairs of pants. In fact, $R$ is
generated by two types of ``4-point'' relations. The first (see figure
\ref{relation1}) corresponds to relations between different trees while the
 second (one of which is depicted in figure \ref{relation2}) corresponds to
relations between the same type of tree but with different homotopy classes
of curves separating the same 4-holed sphere into pairs of pants.

\begingroup
\input{psfig}
\begin{figure}[h]
\centerline{\psfig{figure=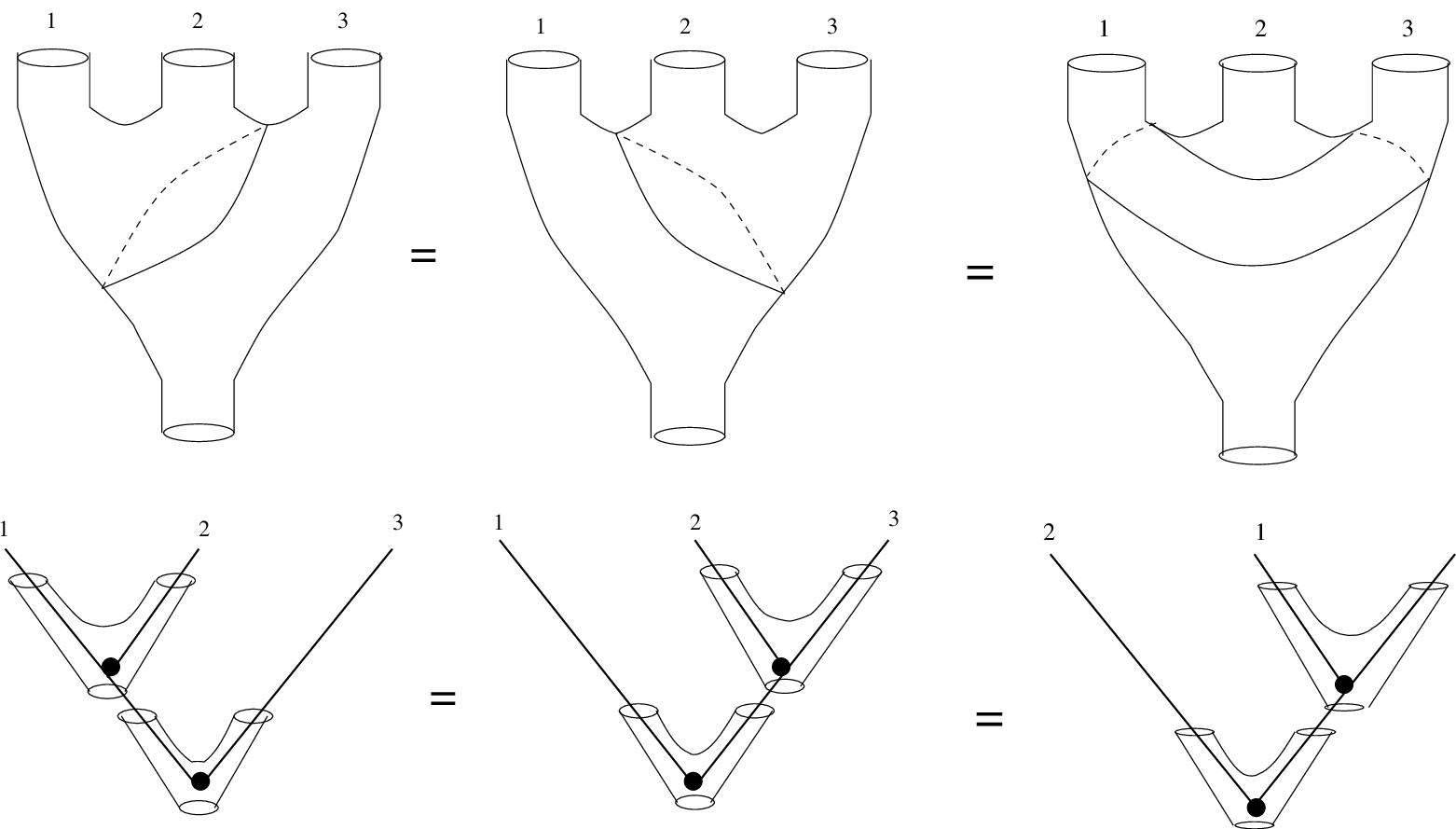,height=2.5in}}
\caption{4-Point Relations Between Different Trees}
\label{relation1}
\end{figure}

\begin{figure}[h]
\centerline{\psfig{figure=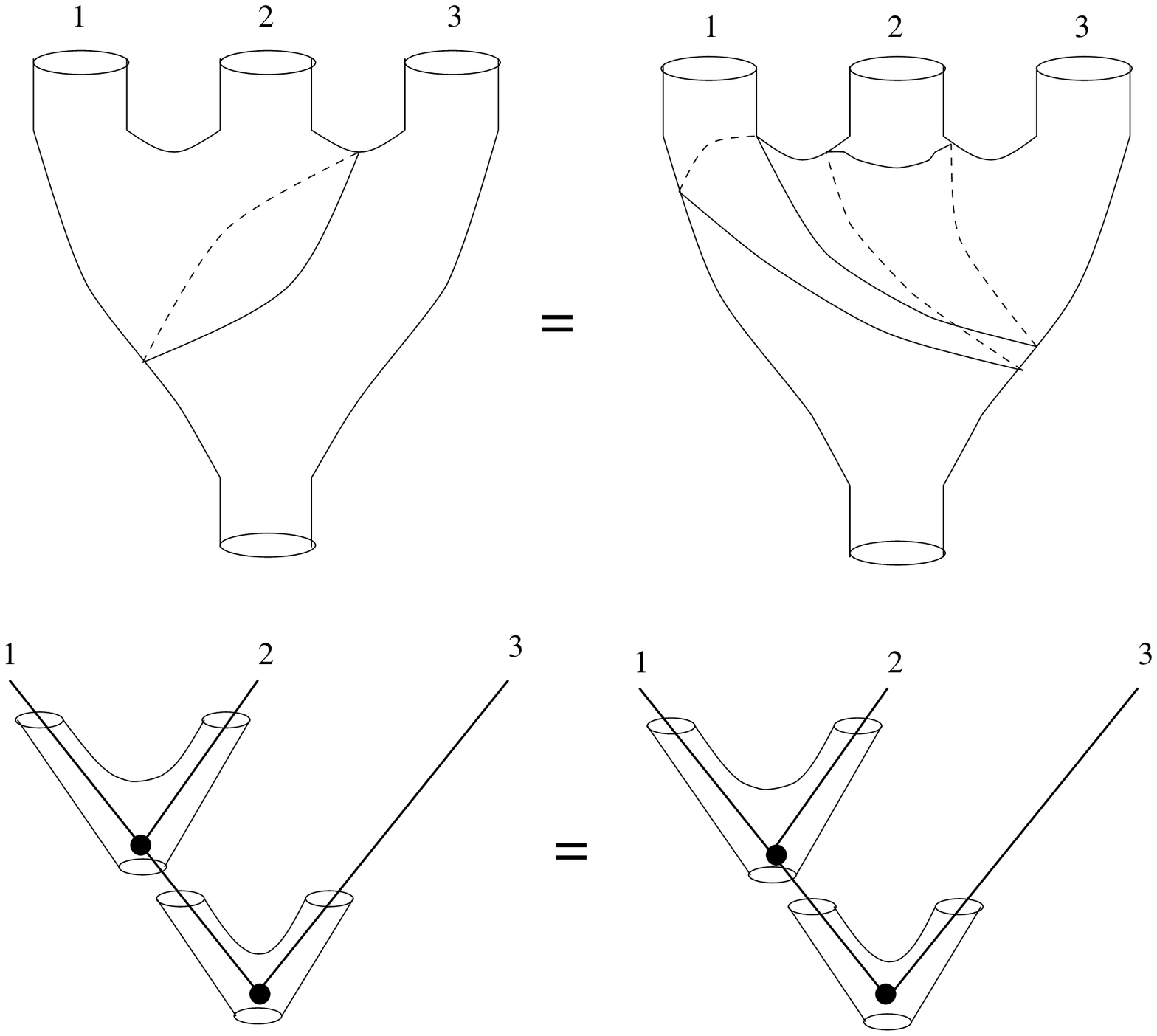,height=2.5in}}
\caption{A 4-Point Relation Between the Same Trees}
\label{relation2}
\end{figure}
\endgroup

\begin{th}[Coherence Theorem]
The operad $\p(n)$ is quadratic, with pairs of pants as generators for $E$ and
the ``four-point'' relations as defining relations.
\end{th}

\begin{pf}
We have a natural surjective morphism of operads
\[
Q(K,E,R) = F(E)/(R) @>>> \p
\]
where $(R)$ is the ideal in $F(E)$ generated by $R$.
To show that this is an isomorphism, we have to show that all
relations between elements of $F(E)(n)$ are consequences of $R$.
Suppose we have a relation between two elements of $F(E)(n)$,
corresponding to binary trees $T_1$ and $T_2$. This means we have a
Riemann sphere with $n+1$ holes and two decompositions into pairs of
pants on it. Applying the four-point identities of figures \ref{relation1}
and \ref{relation2} to fragments of the trees, we can get from $T_1$ to
$T_2$. On the Riemann sphere, we can interpret those moves as erasing the cut
between two subsequent pairs of pants and cutting them along another cut
separating another pair of vertices on the same four-point fragment.
Finally, we arrive at two decompositions into pairs of pants of the
same $n+1$-holed Riemann sphere corresponding to one and the same
tree.

Now, working from the root to the top, we can make the cuts
homotopic if we use the relations \ref{relation2} corresponding to one and
the same tree and different choices of homotopy classes of cuts separating
holes $a$ and $b$. These relations follow from \ref{relation2}.

Thus, we have two decompositions into pairs of pants where cuts are all
homotopic. Finally we can use elements of $K$ to move the cuts along the legs
to get identical decompositions. This proves the theorem.
\end{pf}

\subsection{The Koszul dual $\p^! (n)$}
\label{koszul}

\begin{sloppypar}
The Koszul dual to a quadratic operad $Q(K,E,R)$ is the quadratic
operad $Q( K^{\op}, E^{\vee}, R^\perp)$, cf.\ Ginzburg-Kapranov
\cite{gk}. Here $K^\op$ is the algebra $K$ with the reversed
multiplication. Geometrically, we can regard this as $PK$, where $P$
interchanges the input and output of a cylinder.
\end{sloppypar}

Here is how the Koszul dual is defined. For a $(K, K^n)$-module $V$
with an $S_n$ action, we consider the $(K^\op, (K^\op)^n)$-module
$V^\vee = \Hom_K (V, K)$, $K$ in the subscript acting on $K$ by the
left multiplication and on $V$ via the left action. We will provide
$V^\vee$ with the natural structure of an $S_n$-module, twisted by the
sign representation.

Having our $(K,K^2)$-module $E$ with an $S_2$ action, we take the space
$E^\vee$ as the generating space of the Koszul dual. Using the natural mapping
$F(E^\vee)(3) \to F(E)(3)^\vee$, we can define $R^\perp \subset F(E^\vee)(3)$
as the annihilator of $R \subset F(E)(3)$.

\begin{prop}
\[
R^\perp = \p(3)^\vee,
\]
that is, $R^\perp$ consists of $K$-valued $K$-equivariant functions
$C(S)$ of classes $S$ of $4$-holed spheres.
\end{prop}

Thus, the space $\p^!(n)$ is a quotient of the space of $K$-valued
$K$-equivariant functionals $C(S,\gamma)$ of an isomorphism class $S$
of a Riemann sphere with $n+1$ holes and a homotopy class $\gamma$ of
a decomposition of $S$ into pairs of pants. The space of relations of
this quotient space is the ideal generated by $\p(3)^\vee$.

\begin{prop}
For all $n \ne 1$, $\p^!(n)$ is the quotient space of $K$-valued
$K$-equivariant functions $f(S,\gamma)$, $S \in
\p(n)$, $\gamma$ being a partition of $S$ into pairs of pants
corresponding to a tree diagram, considered up to homotopy, by the
subspace generated by functions $g_T (S,\tilde{\gamma})$, where $S \in
\p(n)$ and $\tilde{\gamma}$ is a partition of $S$ into one $4$-holed
sphere and a few pairs of pants corresponding to a tree $T$ with a
marked $3$-subtree. In particular, $\p^!(1) = \p(1)^\op$, $\p^!(2)
= \p(2)^\vee$ and $\p^!(3) = F(\p(2))(3)^\vee/
\p(3)^\vee$.
\end{prop}

\subsection{The operad cohomology}
\label{operad-coh}

For a quadratic operad $\cal O = Q(K,E,R)$ and an $\cal O$-algebra $A$
and a module $M$ over $A$, define a cochain complex with the $n$th
term
\begin{multline}
\label{chains}
C^n = C^n_{\OO} ( A, M) \\
 =
 \Hom_{K} (F (E^\vee)(n)^\vee \otimes_{S_n, K^n}
A^{\otimes n}, M) / ( \langle R \, \rangle (n) \otimes_{S_n, K^n} A^{\otimes n}
)^\perp, \qquad n \ge 1,
\end{multline}
where, by a certain abuse of notation, $F (E^\vee)(n)^\vee :=
\bigoplus_{\text{binary $n$-trees $T$}} E(T) \otimes \det (T)$, and
$\det (T)$ is the top exterior power of the vector space generated
by all of the internal edges of a tree $T$ (those edges which are not
inputs to $T$). Also, $K$ here acts by the left action on $F
(E^\vee)(n)^\vee$ and $M$.
The subspace $\langle R \rangle (n)
\subset F(E^\vee)(n)^\vee$ is defined as follows:
\[
\langle R \rangle (n) = \bigcap_{\text{1-ternary $n$-trees $T$}} (R \otimes
\det) (T),
\]
where a 1-ternary tree $T$ is a tree with 1 ternary vertex $u$ and
binary vertices $v$ otherwise and
\[
(R \otimes \det) (T) = (R \otimes \det) (\In u) \otimes
\bigotimes_{\text{binary $v \in T$}} E(\In v) \otimes \det (\In v) \otimes \det
\{\text{terminal edges}\},
\]
$ (R \otimes \det ) (\In u) $ being the same as $R$ except that it is
labeled by the incoming edges $\In u$ of the vertex $u$ and that it is twisted
by the $\det$.

\begin{rem}
If $K$ were a field and the space $E$ finite dimensional over $K$, the space
$\langle R \, \rangle$ would be the same as $(R^\perp)^\perp$,
the subspace in $F (E^\vee)^\vee$ which vanishes when contracted with
elements in the operad ideal $(R^\perp) \subset F(E^\vee)$. The following dual
description of the
cochain complex would also be true:
\begin{align*}
C^n
 & \cong  \Hom_{K} (\langle R \rangle (n) \otimes_{S_n, K^n}
A^{\otimes n}, M)\\
 & \cong
\Hom_{S_n, K^n} ( A^{\otimes n}, \operatorname{Sgn}_n \otimes \OO^!(n)
\otimes_{K} M),
\end{align*}
$\operatorname{Sgn}_n$ being the sign representation of $S_n$.
\end{rem}

The differential $d: C^{n-1} \to C^{n}$ can be defined
as follows. A vertex $v$ of a tree $T$ is called {\it extremal},
if all of its incoming edges are terminal for the whole tree. For an
extremal vertex $v$ of a tree $T$, let $T_v$ denote the subtree formed
by the vertex $v$ along with its incoming edges, and $T/v$ the tree
obtained by removing $T_v$ from $T$.  Let $[n]$ denote the set $\{1,\ldots,n
\}$. We label the tree $T/v$ by the set $[n]/v$ which consists of the
elements of $[n]$ which label $T$, adding the label $v$, and omitting the
labels we have just erased. Since we ultimately will consider maps which are
invariant under the symmetric group, it will not matter which
$n-1$-element set we used to label $T/v$. If there is an external edge coming
into the root of the tree, we define a tree $T/r$ by erasing the tree $T_r$
formed by the root vertex and its incoming edges. We label the $T/r$
similarly to $T/v$.

The space $F(E^\vee)(n)^\vee$ is the direct sum of the spaces
$E(T)\otimes \det (T)$ of binary $n$-trees decorated with elements of
$E$ at vertices. The action of $\cal O$ on $A$ defines
\[
E (T_v) \otimes A \otimes A \to A
\]
and thereby
\[
d_v:  \Hom(E(T/v) \otimes \det (T/v) \otimes A^{\otimes [n]/v}, M) \to
\Hom (E(T) \otimes \det (T) \otimes A^{\otimes n} , M),
\]
where for the determinants we used the mapping
\begin{align*}
\det (T) & \to \det (T/v)\\
e_1 \wedge \dots \wedge e_{n-2} & \mapsto e_1 \wedge \dots
\wedge \hat e_k \wedge \dots e_{n-2},
\end{align*}
$e_k$ being the internal edge of $T$ missing from $T/v$.

Similarly, for each external edge $e$ coming into the root, we have our
module structure morphism
\[
E(T_r) \otimes A \otimes M \to M
\]
and
\[
d^e_r : \Hom(E(T/r) \otimes \det (T/r) \otimes A^{\otimes [n]/r}, M)
\to \Hom( E(T) \otimes \det (T) \otimes
A^{\otimes n}, M) .
\]
If there are no external edges coming into the root of $T$, we put
$d^e_r=0$. For a binary tree $T$, two external edges may come into the
root, if and only if $T$ is a 2-tree. This is the only case when there is
more than one mapping $d^e_r$.

Thus we can form a mapping $d= \sum_v d_v +
\sum_e d^e_r$
\[
d: \Hom_{K} ( F (E^\vee)(n-1)^\vee
\otimes_{S_{n-1}, K^{n-1}} A^{\otimes n-1}, M) \to
\Hom_{K} (F (E^\vee)(n)^\vee \otimes_{S_n, K^n} A^{\otimes n}, M).
\]

The differential $d: C^1 = \Hom_K (A, M)  \to C^2= \Hom_K (E \otimes_{S_2, K}
A \otimes A , M)$ may need special attention, as it involves cutting
cylinders off a pair of pants. If $c: A \to M$ is a 1-cochain,
then
\[
(dc)(S; a_1, a_2) = X(S; a_1, c (a_2) ) - c (Y(S; a_1, a_2)) + X(\tau S; a_2,
c(a_1)),
\]
where $S \in E$, $a_1$, $a_2 \in A$, $X: E \otimes A \otimes M \to M$ is the
module structure morphism, $Y: E \otimes A \otimes A \to A$ is the algebra
structure morphism, and $\tau \in S_2$ is the transposition.

\begin{lm}
\label{d^2}
The subspaces
\[
(\langle R \, \rangle (n) \otimes_{K^n} A^{\otimes n})^\perp
\]
are preserved by $d$ and the resulting map on the quotients $C^n$
is a differential, i.e., $d^2 = 0$.
\end{lm}

\begin{pf}
If we have a cochain $c$ in $C^n$ which vanishes on $(R \otimes \det) (T)$
for a 1-ternary tree $T$, terms of its differential $dc$ will vanish on $(R
\otimes \det) (T')$ for 1-ternary $n+1$-trees $T'$ obtained from $T$ by
grafting binary trees onto it.  Therefore, they will vanish on $\langle R
\rangle (n+1)$ as well.

For a cochain $c \in C^n$, those terms of $d^2 c$ which are obtained by
erasing the vertices of a 1-ternary tree $T$ whose ternary vertex is terminal
will vanish on $(R \otimes \det) (T)$, because $R$ as the relation spaces
acts on $A^3$ by zero. There are also terms in $d^2 c$ corresponding to the
root vertex - they will vanish on $(R \otimes \det) (T)$ corresponding to a
1-ternary tree $T$ whose ternary vertex is the root vertex, because $R$ acts
by zero on $A^2 \otimes M$. Terms of $d^2 c$ whose terms are too far away to
be united in a 3-subtree, will cancel each other because of the signs imposed
by $\det (T)$.
\end{pf}

\begin{df}
The {\it operad cohomology\/} $H_{\OO}^n (A, M)$, $n \ge 1$, of an
algebra $A$ over an operad $\OO$ with values in a module $M$ is the
cohomology of the complex $C_{\OO}^n (A, M)$.
\end{df}

\begin{rem}
\label{classical}
When $\OO = \As$, we obtain the Hochschild complex
\[
\Hom (A, M) \to \Hom (A^2, M) \to \dots,
\]
when $\OO = \Lie$, we obtain the standard complex
\[
\Hom (A, M) \to \Hom (\Lambda^2 A, M) \to \dots,
\]
when $\OO = \Comm$, we obtain the Harrison complex
\[
\Hom (A, M) \to \Hom (S^2 A, M) \to \dots.
\]
Thus the operad cohomology in these cases is Hochschild, Lie algebra
and Harrison, respectively. We expect any other classical and new
examples to fit this scheme. For instance, applying the above to the
noncommutative Poisson operad $\OO = \cal Poiss$, we get the complex
giving the cohomology of a noncommutative Poisson algebra, see
\cite{fgv}. These complexes are missing their zeroth terms, which will be
added in the following section.
\end{rem}

\subsection{Digression: the zeroth term}

This section is a brief discussion how to modify the formalism of quadratic
operads, so as to include the zeroth term in the cochain complex. It is in
fact digression from our main operad $\p$ of moduli spaces, because it is not
symmetric in the following sense.

A {\it symmetric quadratic operad with vacuum\/} $\OO = Q(C,K,E,R)$ is
a quadratic operad with $\OO(1)= K$, the space $E$ of generators and
the space $R \subset F(E)(3)$ of defining relations to which a {\it
vacuum space\/} $\OO(0) = C$ is added. The composition maps are
extended so that they satisfy the associativity axioms and the
equivariance under the action of the permutation groups. The word
symmetric menas that in addition, {\it the two right actions $\oo_1$ and
$\oo_2$ of $K$ on $E$ must coincide}.

We also need the following data and axioms, satisfied in all examples of
Remark~\ref{classical}, see the examples below.

\begin{quote}
A morphism $\phi = \oo_1: E \otimes_K C \to K$ of $(K,K)$-modules,
$K$ acting on $E$ via the right module structure, along with a
right inverse $s: K \to E \otimes_K C$, $\phi s = \id_K$.
\end{quote}

Here is the axiom relating the above data associated with $C$ to
$R$. {\it We require that the image of the composition of certain maps
\begin{equation}
\label{axiom}
E @>\delta_1>> (E \otimes_K K)^{\oplus 3} @>\delta_0>> F(E)(3) \otimes_K C
\end{equation}
be contained in $R\otimes_K C$}. Here we should think of the three
components of the middle term as being indexed by the three trees
shown in figure \ref{threeterms}, all of which are allowed to have
vertices of valence 1.
\begingroup
\input{psfig}
\begin{figure}[h]
\centerline{\psfig{figure=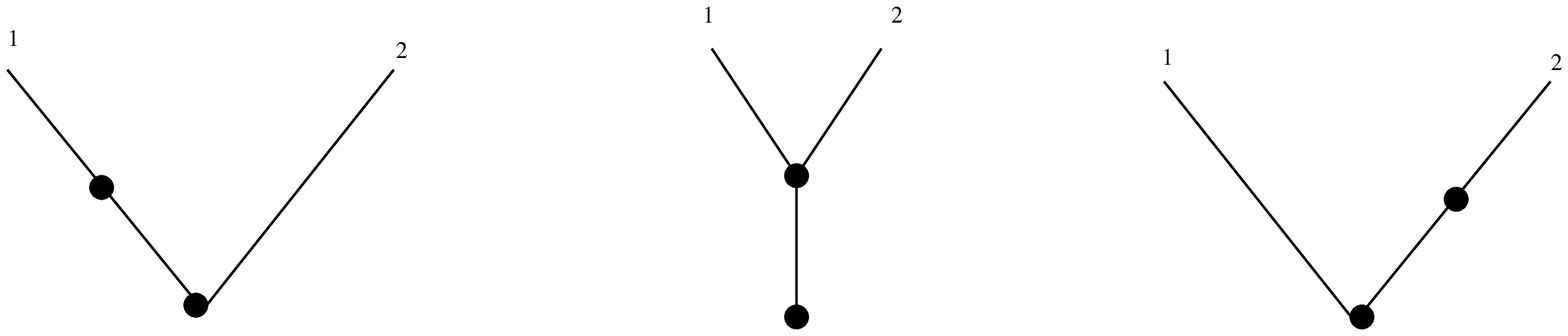,height=1in}}
\caption{Trees Indexing $(E \otimes_K K)^{\oplus 3}$}
\label{threeterms}
\end{figure}
\endgroup
The mappings $\delta_1$ and $\delta_0$ are defined as follows:
\[
\delta_1 (e) = (e \otimes \id)_1 - (e \otimes \id)_2 + (e \otimes \id)_3 ,
\]
the outer
indices refer to the component of $(E \otimes_K K)^{\oplus 3}$ (see
figure \ref{d1}).
\begingroup
\input{psfig}
\begin{figure}[h]
\centerline{\psfig{figure=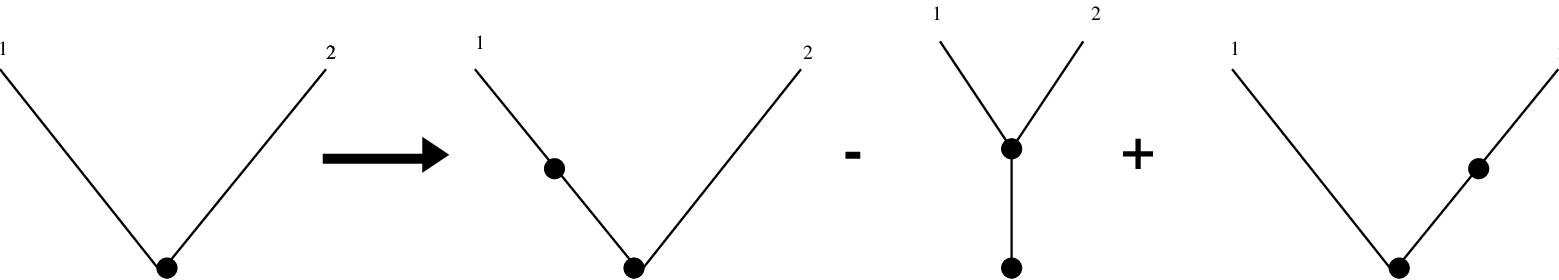,height=1in}}
\caption{Action of $\delta_1$}
\label{d1}
\end{figure}
\endgroup
The action of $\delta_0$ on a component of $( E \otimes_K K)^{\oplus 3}$
corresponding to a tree $T$ is equal to
\[
\delta_0 = \id_E \otimes s -\tau \otimes s
\]
where $s$ acts by replacing the 1-corolla (the subtree consisting of a vertex
with exactly one incoming edge) by a difference of two trees as shown in figure
\ref{s} and $\tau$ is the transposition  in $S_2$ acting on $E$. The assumption
that the two right actions of $K$ on $E$ are equal makes the second term of
$\delta_0$ correctly defined.
\begingroup
\input{psfig}
\begin{figure}[h]
\centerline{\psfig{figure=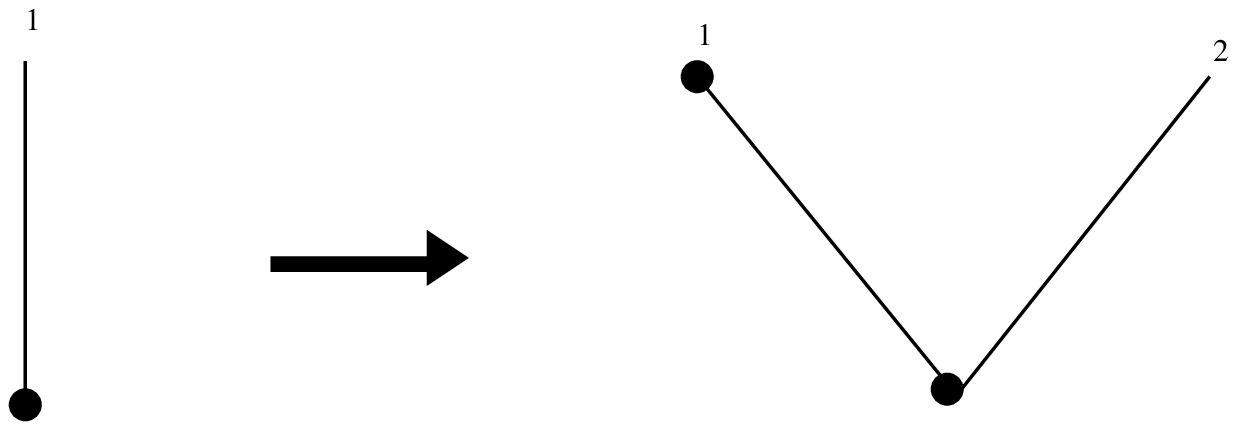,height=1in}}
\caption{Action of $s:K\,\to\,E\otimes_K C$}
\label{s}
\end{figure}
\endgroup
In particular, $\delta_0$ acts on the first component of  $( E \otimes_K
K)^{\oplus 3}$ as shown in figure \ref{d0} and similarly for the other two
components.
\begingroup
\input{psfig}
\begin{figure}[h]
\centerline{\psfig{figure=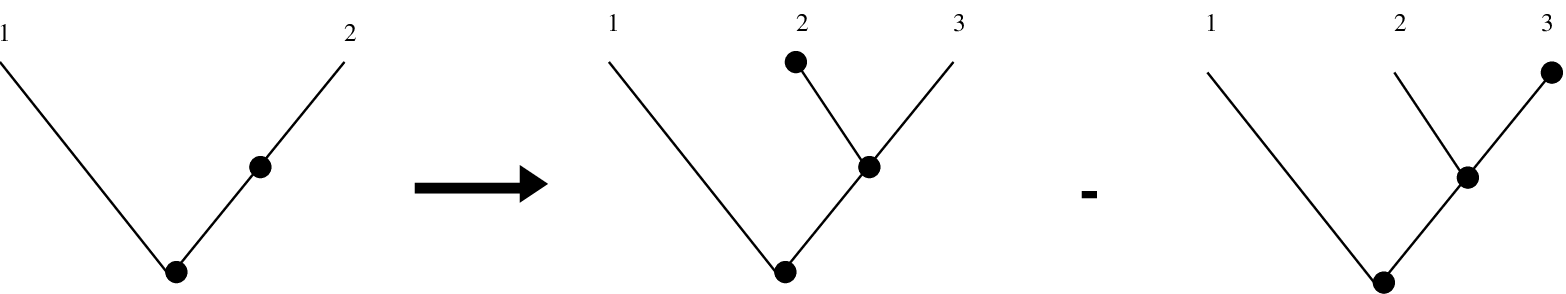,height=1in}}
\caption{Action of $\delta_0$ on a Component of
$(E \otimes_K K)^{\oplus 3}$}
\label{d0}
\end{figure}
\endgroup

An algebra over such an operad $\cal O(n)$ is defined as a vector
space $V$ along with a morphism $\{\cal O(n) \to \End{V}(n) \; | \; n
\ge 0\}$ respecting all of the data.

Let us illustrate the above abstract nonsense with the examples of the
standard algebraic operads $\As$, $\Lie$ and $\Comm$, describing
associative, Lie and commutative algebras.

\begin{sloppypar}
\begin{ex}[The associative operad $\As$]
\label{assoc}
Consider the alphabet
$\{x_1, x_2, \dots\}$ and words constructed from it. Then $C$ is
generated by the empty word $()$ and $K$ by the word $x_1$ - both spaces are
one-dimensional. $E$ is generated by two words $x_1 x_2 $ and $x_2
x_1$, $F(E)(3)$ is generated by all possible permutations of the words
$(x_1 x_2) x_3$ and $x_1(x_2 x_3)$ and $R$ is generated by 6
associators of the kind $(x_1 x_2) x_3 - x_1(x_2 x_3)$. The mappings
$\psi$ and $\eta_{1,2,3}$ are obvious, because we have the identity
element $\id = x_1 \in K$, the mapping $\phi$ takes both $x_1 x_2
\otimes ()$ and $x_2 x_1 \otimes ()$ to $x_1$, and $s: x_1 \mapsto x_1
x_2 \otimes ()$. The interested reader may check that $\delta_0
\delta_1$ maps $x_1 x_2$ to the linear combination of three associators:
$(x_1 x_3) x_2 - x_1 (x_3 x_2) - ((x_3 x_1) x_2 - x_3 (x_1 x_2)) - (
(x_1 x_2) x_3 -x_1 (x_2 x_3)) \in R$, tensored with $()$.
\end{ex}
\end{sloppypar}

\begin{ex}[The Lie operad $\Lie$]
In this case $C$ is generated by $[\,]$, $K$ by
$x_1$, $E$ by $[x_1, x_2]$ and $R$ by the Jacobi identity, see \cite{gk} for
more details. Since $C$, $K$ and $E$ are naturally identified with complex
numbers $\nc$, we can define all mappings $\phi$, $s$, $\psi $ and
$\eta$'s as $\id: \nc \to \nc$. Then $\delta_0 \delta _1$ maps $[x_1 x_2]$
to twice the Jacobi identity, tensored with $[\,]$.
\end{ex}

\begin{ex}[The commutative operad $\Comm$]
This case is similar to $\Lie$, but here $\delta_0$ is already zero,
because $x_1 x_2 = x_2 x_1$.
\end{ex}

Now we are up to add the zeroth term to the complex $C^\bullet =
C_{\OO}^\bullet (A,M)$. By definition,
\[
C^0 = C^0_{\OO} (A, M) = \Hom_K (C, M).
\]
The differential
\[
d: C^0 = \Hom_K (C, M) \to C^1 = \Hom_K (A, M) ,
\]
is defined similar to all higher differentials above. There are no
terms $d_v$ corresponding to extremal vertices, but there are two
terms $d_r^1$ and $d_r^2$ corresponding to the root. The term $d_r^1$ is the
composition of
\[
\Hom (C, M) \to \Hom( E  \otimes_K C\otimes_K A , M),
\]
which is induced by the structure morphism
\begin{equation}
\label{module}
E \otimes A \otimes M \to M,
\end{equation}
and
\[
\Hom( (E  \otimes_K C)\otimes_K A , M) \to \Hom( K \otimes_K A , M),
\]
which is induced by the structure morphism
\begin{equation}
\label{cut}
s: K \to E \otimes_K C .
\end{equation}
The term $d_r^2$ is merely $(d_r^1)^\tau$, where we compose $s$ with the
transposition $\tau \in S_2$ acting on $E$.

The axiom \eqref{axiom} ensures that $d^2 = 0$, as in
Lemma~\ref{d^2}. Thus we now have a complex
\[
0 \to C_{\OO}^0 (A, M) \to C_{\OO}^1 (A, M) \to C_{\OO}^2 (A, M) \to \dots ,
\]
whose cohomology groups $H_{\OO}^n (A, M)$, $n \ge 0$, form the {\it
cohomology of an algebra over a symmetric operad with vacuum}.

\begin{rem}
In the above cases of operads $\OO = \As$, $\Lie$ and $\Comm$, adding
the zeroth term to the complex $C_{\OO}^\bullet (A, M)$ completes the
complexes of Remark~\ref{classical} by including the standard zeroth
terms. For the Harrison complex, the operator $d: C^0 \to C^1$ will be
equal to zero, as usual.
\end{rem}

\subsection{Description of the complex for the moduli space operad}

Here we are going to apply the construction of the previous sections to the
operad $\p(n)$, $n \ge 1$, of isomorphism classes of Riemann spheres with $n+1$
labeled holomorphically embedded disks.

Suppose we have a (tree level $c=0$) conformal field theory (CFT),
which is nothing but an algebra $A$ over the operad $\p(n)$, and a
module $M$ over this algebra. One may require these algebras and modules be
smooth, i.e., the action of the operad $\p(n)$ depends smoothly on the point
in $\p(n)$, which may be easily included in our formalism. We do not require
smoothness for two reasons: the noncontinuous cohomology of Lie groups
makes perfect sense and the action of the Virasoro algebra, which
is assumed implicitly through the action of $K = \p(1)$, may be regarded
as a kind of smoothness condition - $\p(n)$ is a locally homogeneous
space over $n+1$ copies of the Virasoro algebra, which acts through
$K^{n+1}$, and the equivariance of algebra and module structure maps
as well as cochains with respect to $K^{n+1}$ will give rise to a
stronger condition than just smoothness.

We may also consider holomorphic algebras and modules over $\p(n)$ and
holomorphic cochains, which brings us to the case of chiral or vertex
operator algebras and modules over them. This is one of the main
points of the second part of the paper.

We will consider cochains \eqref{chains}, and our
description of cochains will be similar to that of the Koszul dual
operad $\p^! (n)$ in Section~\ref{koszul}.

 The space $C_{\p}^n (A, M)$ of cochains of a CFT, $n \ge 1$, is a
quotient space of the space $ \Hom_{K} (F (E^\vee)(n)^\vee
\otimes_{S_n, K^n} A^{\otimes n}, M)$ of $K$-equivariant
functions
\begin{equation}
\label{f}
f(S, \gamma, e_T; a_1, a_2, \dots , a_n)
\end{equation}
where $S \in \p(n)$, $\gamma$ is a partition of $S$ into pairs of pants
corresponding to a tree diagram $T$, considered up to homotopy, $e_T = e_1
\wedge \dots \wedge e_{n-2} \in \det (T)$, $e_1, \dots , e_{n-2}$ being
internal edges of $T$, and $a_1, \dots, a_n \in A$. Including $e_T$ as a
variable may be interpreted as choosing an orientation on the space
generated by internal edges. The functions $f(\sigma S, \gamma,
e_T; \linebreak[0] a_1, a_2, \dots, a_n )$ and $f(S, \gamma, e_T;
a_{\sigma(1)},  a_{\sigma(2)}, \dots, a_{\sigma(n)})$, where $\sigma$ is a
permutation acting on $S$ by relabeling the inputs, are said to be
equivalent. Functions of the form
\[
f((S, \gamma, e_T) \oo_i x; a_1, \dots , a_i, \dots, a_n) \text{ and } f(S,
\gamma , e_T; a_1, \dots , x a_i, \dots, a_n)
\]
for $x \in K$ are also regarded as equivalent. The subspace by which we mod
out is the sum of the following subspaces each of which is
indexed by a 1-ternary $n$-tree $\tilde T$, i.e., a tree whose vertices are
 all binary with the exception of one which is ternary.

The subspace corresponding to a 1-ternary tree $\tilde T$ is the space of
functions
\[
g(S, \tilde \gamma, e_{\tilde T}; a_1, a_2, \dots , a_n) \in M,
\]
where $\tilde \gamma$ is a partition of $S$ into one $4$-holed sphere and
some number of pairs of pants corresponding to the tree $\tilde T$. Such
functions form a subspace of $ \Hom_{K} (F (E^\vee)(n)^\vee \linebreak[0]
\otimes_{S_n,K^n}
A^{\otimes n}, M)$, because if $\gamma$ is a partition of  $S$ into pairs of
pants, which coincides with $\tilde \gamma$ after erasing one cut, then we
 set $g(S, \gamma, e_T; \dots) = g(S, \tilde \gamma, e_{\tilde T}; \dots)$,
where $e_T = e_{\tilde T} \wedge e_s$, $s$ being the edge of $T$ missing in
$\tilde T$. We set  $g(S, \gamma , e_T; \dots) = 0$ otherwise.

The terms $C^1$ and $C^2$ consist of functions
\[
f(S ; a_1, \dots , a_n) \in M,
\]
where $S$ is a cylinder when $n=1$ and a pair of
pants when $n=2$. For $n=3$, $C^3$ is the quotient of the space of
functions \eqref{f} by the subspace of functions which are independent
of the partition into pairs of pants.

The differential acts as follows. Denote by $Y(S; a_1, a_2)$ the $\p$-algebra
structure
\[
Y: E \otimes A \otimes A \to A
\]
on $A$ and by $X(S, a, m)$ the $A$-module structure
\[
X: E \otimes A \otimes M \to M
\]
on M.
For an $n$-cochain $f$, as in \eqref{f}, $n\ge 2$,
\begin{multline*}
(df) (S, \gamma, e_T; a_1, a_2, \dots , a_{n+1}) \\
\begin{split}
=& \sum_{\text{extremal vertices $v$}} f(S_1^v, \gamma^v, e_{T/v} ; Y(S_2^v;
a_i, a_j), a_1, \dots, \hat a_i, \dots , \hat a_j, \dots, a_{n+1}) \\
& + \sum_{\left\{\parbox{.72in}{\scriptsize terminal edges $e$ pointing to the
root}\right\}} X (S_1^e; a_i, f(S_2^e, \gamma^e, e_{T/r} ; a_1, \dots , \hat
a_i, \dots. a_{n+1})),
\end{split}
\end{multline*}
where $(S, \gamma)$ is an $n+2$-holed Riemann sphere partitioned into pairs of
pants, the first summation runs over extremal vertices $v$ of the diagram $(S,
\gamma)$, $i$ and $j$ are the labels of the vertices right above $v$, $S_2^v$
is the pair of pants corresponding to $v$, $(S_1^v, \gamma^v)$ is what remains
of $(S, \gamma)$ after removing $S_2^v$, and $T/v$ is the tree corresponding to
$(S_1^v, \gamma^v)$. The holes of $S_1^v$ are labeled as follows. The hole left
after removing $S_2^v$ goes first, then all others following their order on
$S$. $e_{T/v}$ is just the wedge product of the internal edges of $T/v$ in the
same order they were in $e_T$. The holes of $S_2^v$ labeled with $i$ on $S$ is
labeled by 1, the other input hole by 2. Similarly, the second summation runs
over terminal edges $e$ going into the root of $(S, \gamma)$, if any, $i$ is
the label at the input of the edge $e$, $S_1^e$ is the pair of pants
corresponding to the root and $(S_2^e, \gamma^e)$ is what remains after
removing $S_1^e$ from $(S, \gamma)$. Also, $T/r$ is the tree corresponding to
$(S_2^e, \gamma^e)$, the holes of $S_2^e$ are labeled in the same order they
were on $S$, the hole labeled with $i$ on $S$ is labeled with 1 on $S_1^e$, and
$e_{T/r}$ is the product of the corresponding edges of $T$ in the same the
order.

The differential for 1-cochains is defined as at the end of
Section~\ref{operad-coh}. Lemma~\ref{d^2} guarantees that the differential $d$
is well-defined and that $d^2 = 0$. This explicit description wil be helpful
when we study deformations of CFT's.

\subsection{The cohomology of TFT's and the Harrison cohomology }

A (tree level) {\it
topological field theory\/} (TFT) is a $\p$-algebra $A$, such that the
structure morphisms $\p (n) \to \End{A}$ are constant. This implies that the
Virasoro semigroup $\p(1)$ acts trivially on $A$ and that the action of a
pair of pants from $\p(2)$ defines the structure of a commutative algebra on
$A$. In this particular case, the above complex $C_{\p}^\bullet (A, M)$ turns
into the Harrison complex $C_{\operatorname Harr}^\bullet (A,M)$ of a
commutative algebra $A$ with values in a module $M$. The second cohomology
group $\operatorname{Harr}^2 (A, A)$ parametrizes classes of infinitesimal
deformations of $A$ within the category of commutative algebras. From
Theorem~\ref{deform} below, it will follow that the same group parameterizes
deformations of a TFT, as well.

\subsection{Deformations of conformal field theories}

We will now show that the complex $C_{\p}(A,A)$ governs the
deformations of $\p$-algebras. Let $A$ be a $\p$-algebra where $\p =
\{\,\p(n)\,\}_{n\geq 1}$ and the morphism is denoted by
$Y:\p\,\to\,\End{A}$. Since $\p$ is a quadratic operad, the algebra
structure is completely determined by $K$-module structure of $A$ and
the action of $E$ on $A$. Let $A$ be a $\p$-algebra
then $A$ is a $K$-module. The problem is to classify inequivalent
infinitesimal deformations of $A$ as a $\p$-algebra keeping the $K$-module
structure fixed. An {\it infinitesimal deformation of a CFT $A$\/} is the
structure of a $\p$-algebra on $A \otimes \nc [t]/(t^2)$ over the
algebra $\nc [t] /(t^2)$ of {\it dual numbers}, such that the
reduction of this structure at $t=0$ coincides with the initial
$\p$-algebra structure on $A$ over $\nc$. Two infinitesimal
deformations are said to be {\it equivalent}, if there is an isomorphism
between the two $\p$-algebras over $\nc [t]/(t^2)$ inducing the
identity morphism at $t=0$. {\it Keeping the $K$-module structure on
$A$ fixed\/} means that we assume this structure is extended trivially to
$A \otimes \nc[t]/(t^2)$.

Infinitesimally, a $K$-module structure on $A$ is the structure of a
module over the Virasoro algebra $\operatorname{Lie} (K)$. Classes of
infinitesimal deformations of the $K$-module structure on $A$ are
parameterized by the Lie algebra cohomology $H^1(\operatorname{Lie}
(K); \operatorname{End} (A))$. Writing down conditions for a
deformation of a CFT, where the action of $K$ is allowed to vary, is an
easy exercise. However, its cohomological interpretation is not obvious and
must combine the operad cohomology and the Lie algebra cohomology in a new
kind of spectral sequence.

It is worth pointing out that for the arguments in this section, it
makes no difference whether $\p$ is the moduli space of Riemann
spheres with holomorphically embedded disks which are allowed to
intersect at their boundaries or whether $\p$ is the same but where
the disks are not allowed to intersect at all.

The relations $R$ give rise to an associativity condition on the binary
operations on $A$. Let $S$ be a 4-holed sphere in $\p(3)$ with any two
partitions of $S$, $\gamma$ and $\gamma',$ into two pairs of pants in $\p(2)$
then the associativity relations are
\begin{equation*}
(Y \oo Y)_{(S,\gamma)} = (Y\oo Y)_{(S,\gamma')}
\end{equation*}
where $Y :  \p(2)\,\to\,\End{A}(2)$ is the algebra map and $(Y \oo Y)_{(S,
\gamma)} $ denotes the composition indicated by the tree associated to
$(S,\gamma)$. Let $Y' = Y + t \alpha$ be a deformation of $Y$ where $\alpha$
is an $S_2$-equivariant and $(K,K^2)$-equivariant (smooth) map $\p(2)\,\to\,
\End{A}(2).$ Notice that $\alpha$ can be interpreted as an element in
$C_{\p}^2 = \Hom_{K}(P(2)\otimes_{S_2, K^2} A^2, A).$ Keeping terms of first
order in $t$, we obtain
\begin{equation*}
(Y \oo \alpha + \alpha \oo Y)_{(S,\gamma)} = (Y \oo \alpha + \alpha \oo
Y)_{(S,\gamma')}
\end{equation*}
which we can rewrite as
\begin{equation*}
(Y \oo \alpha + \alpha \oo Y)_{(S,\gamma-\gamma')} = 0.
\end{equation*}
This means that
\begin{equation*}
(d\alpha)_r := (Y\oo\alpha +\alpha\oo Y)_r = 0
\end{equation*}
for any $r$ in $R$. Interpreting $d\alpha$ as an element of
$\Hom_{K}(F(E)(3)\otimes_{S_3, K^3} A^{\otimes 3}, A)$ the last equation
means that $d\alpha$ belongs to $(R\otimes_{K^3} A^{\otimes 3})^\perp$ and,
therefore, vanishes in $C_{\p}^3(A,A)$. In other words, $\alpha$ is a cocycle
in $C_{\p}^2(A,A)$.

However, it may be that the deformation given above is trivial. This means that
there exists an isomorphism $\Phi = \id_A + t \beta :A\,\to\,A$, $\beta \in
\Hom_K (A, A) = C^1_{\p} (A, A)$, satisfying
\begin{equation*}
\Phi \oo Y' = Y \oo (\Phi \otimes \Phi).
\end{equation*}
Plugging in $Y' = Y+t\alpha$ and we obtain for all $S$ in
$E$ and $v_1,v_2$ in $A$
\begin{equation*}
\alpha(S; v_1, v_2) = Y(S; v_1 , \beta(v_2)) - \beta(Y(S; v_1,
 v_2)) + Y(S; \beta(v_1) , v_2)
\end{equation*}
but the right hand side is nothing more than $d\beta$. We have just shown
the following.
\begin{th}
\label{deform}
The equivalence classes of infinitesimal deformations of a (tree level $c=0$)
conformal field theory (or $\p$-algebra) $A$ which leave the action of the
Virasoro semigroup fixed are classified by elements in $H_{\p}^2(A,A).$
\end{th}

\begin{rem}
In accordance with the general philosophy behind deformation theory, cf.\
Deligne \cite{dl2,dl}, Drinfeld \cite{dr}, Goldman-Millson \cite{gm},
Kontsevich \cite{kon:def} and Schlessinger-Stasheff \cite{ss}, the complex
$C^\bullet_{\p} (A, A)$, governing deformations of a $\p$-algebra $A$ has a
natural structure of a differential graded Lie algebra. This is because
$C^\bullet_{\p} (A, A)$ is obviously an operad, and the compositions
$\circ_i$ on it define a Lie bracket by the formula:
\begin{align*}
[f , g] & = f \circ g \pm g \circ f,\\
f \circ g & = \sum_i \pm f \circ_i g.
\end{align*}
The signs are almost impossible to figure out, as usual. It is better to refer
to \cite{markl,gv}, where this kind of structure for an operad has been
observed. Moreover, results of \cite{gv} imply a richer structure of a brace
algebra on $C^\bullet_{\p} (A, A)$. Another way to see this bracket, as
suggested to us by Stasheff, is to generalize the results of his paper
\cite{jim:intrinsic} to the operadic setting. This means to construct a
canonical isomorphism $C^\bullet_{\p} (A, A) \to \operatorname{Coder}(BA)$,
where $BA$ is the bar construction for the $\p$-algebra $A$, and to carry the
commutator of coderivations over to the cochain complex.
\end{rem}

\section{Deformations and the Cohomology of Algebras over the Little
Intervals Operad and Vertex Operator Algebras}

Having shown that the cohomology theory associated to $\p$-algebras
governs deformations of $\p$-algebras, we shall now do the reverse --  we
shall obtain the cohomology of algebras over the little intervals operad by
studying its deformations mimicking the construction of Hochschild cohomology
from deformations of associative algebras. We begin by realizing the operad
governing associative algebras as a certain moduli space of configurations of
embedded intervals in the oriented circle and then realize the little
intervals operad as a suboperad of a real analog of the operad governing
conformal field theory. In the context of vertex operator algebras, this
leads us to a complex which closely resembles Hochschild cohomology.

{\sl For simplicity, the operads we will consider henceforth will be operads
with unit of the form $\OO = \{\,\OO(n)\,\}_{n\geq 1}$} although many of
these arguments extend to the general case.

\subsection{The associative operad as a one dimensional topological field
theory}

{\sl Henceforth, let $\nr$ be the real line oriented so that the canonical
inclusion $\nr\,\into\,\nc$ is orientation preserving and let $I$ be the
oriented interval $\left[-1,1\right]$.} Consider the oriented circle
$S^1$. For all $n \geq 1,$ let $\A(n)$ be the space of configurations of
$(n+1)$ distinct, ordered, orientation preserving embeddings of $I$ into
$S^1$ where any two such configurations are identified if they are related by
an orientation preserving diffeomorphism of $S^1$.  $S_n$ acts upon $\A(n)$
simply transitively by permuting the ordering of the first $n$ maps. There
are composition maps $\A(n)\times\A(n') \,\to\, \A(n+n'-1)$ taking $(\Sigma,
\Sigma') \,\mapsto\,\Sigma\oo_i\Sigma'$ given by cutting out the $(n'+1)$st
interval of $\Sigma'$ and the $i$th interval of $\Sigma$ and then gluing them
together along their boundaries in an orientation preserving way. These
operations make $\A = \{\,\A(n)\,\}_{n\geq 1}$ into an operad. The
orientation on $S^1$ picks out a canonical element $\e{n}$ in $\A(n)$ for
each $n$ which is the unique class consisting of maps whose ordering
increases as one proceeds around $S^1$ as prescribed by its orientation.
(See figure~\ref{operad}.) This element allows us to identify $\A(n)$ with
$S_n$

\begingroup
\input{psfig}
\begin{figure}[h]
\centerline{\psfig{figure=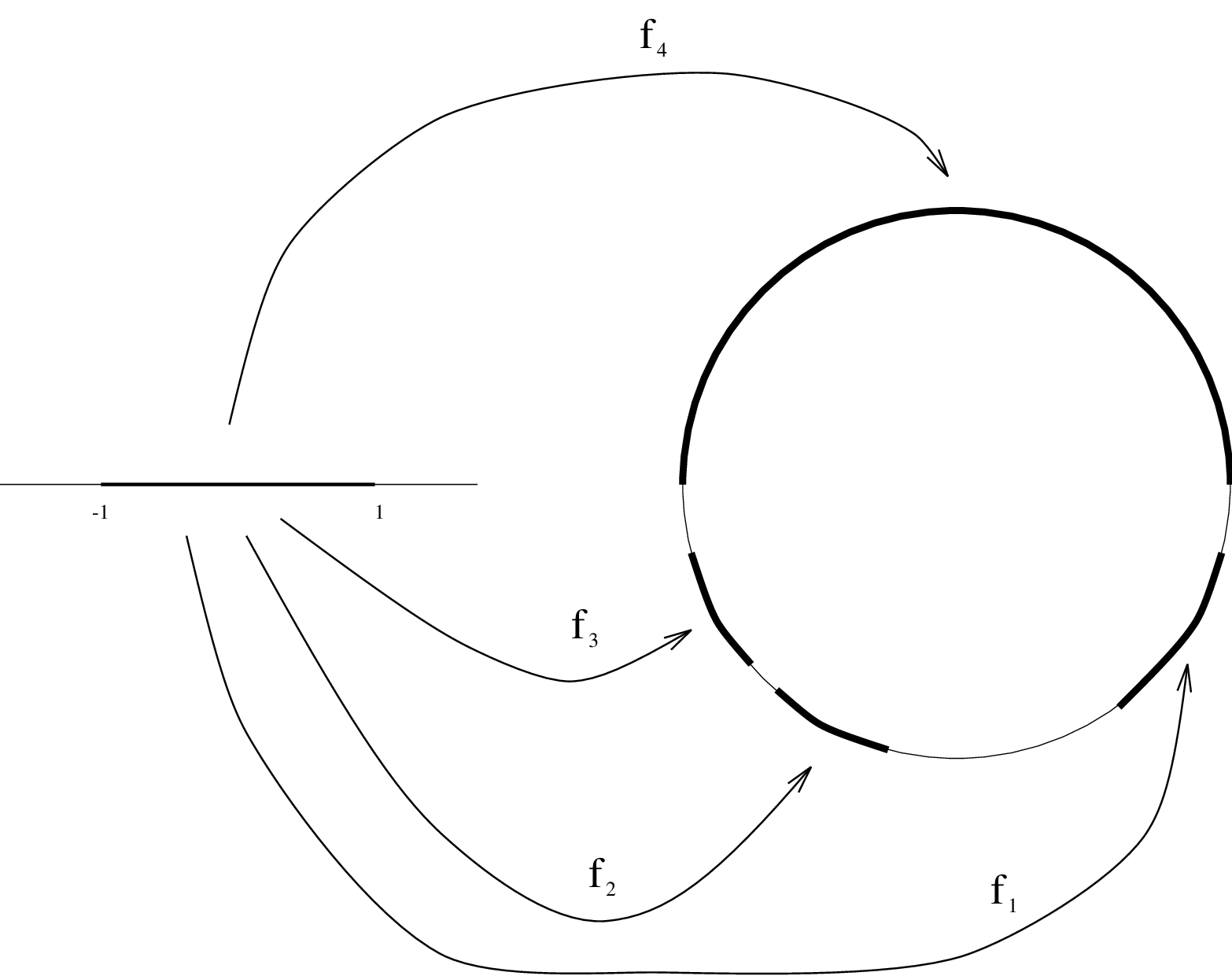,height=3in}}
\caption{The Canonical Element in \A(3)}
\label{operad}
\end{figure}
\endgroup

An $\A$-algebra is nothing more than an associative algebra, hence, $\A$ may
be called the {\sl associative operad}. $\A$ generates an operad of vector
spaces $\As = \{\,\As(n)\,\}_{n\geq 1}$ which is the operad that is usually
referred to as the associative operad, in the literature.

$\As$ is a quadratic operad generated by $(\As(1), \As(2))$ with relations
\begin{equation}
\e{3}\,=\,\e{2}\oo_1\e{2} = \e{2}\oo_2\e{2}\label{eq:assoc}
\end{equation}
as well as similar relations which arise from the action of $S_3$ on this
equation (see figure \ref{associativity}).  This means that all of the
operations of an $\As$-algebra can be obtained from the identity map,
compositions of binary multiplications and the action of the permutation
group.

\begingroup
\input{psfig}
\begin{figure}[h]
\centerline{\psfig{figure=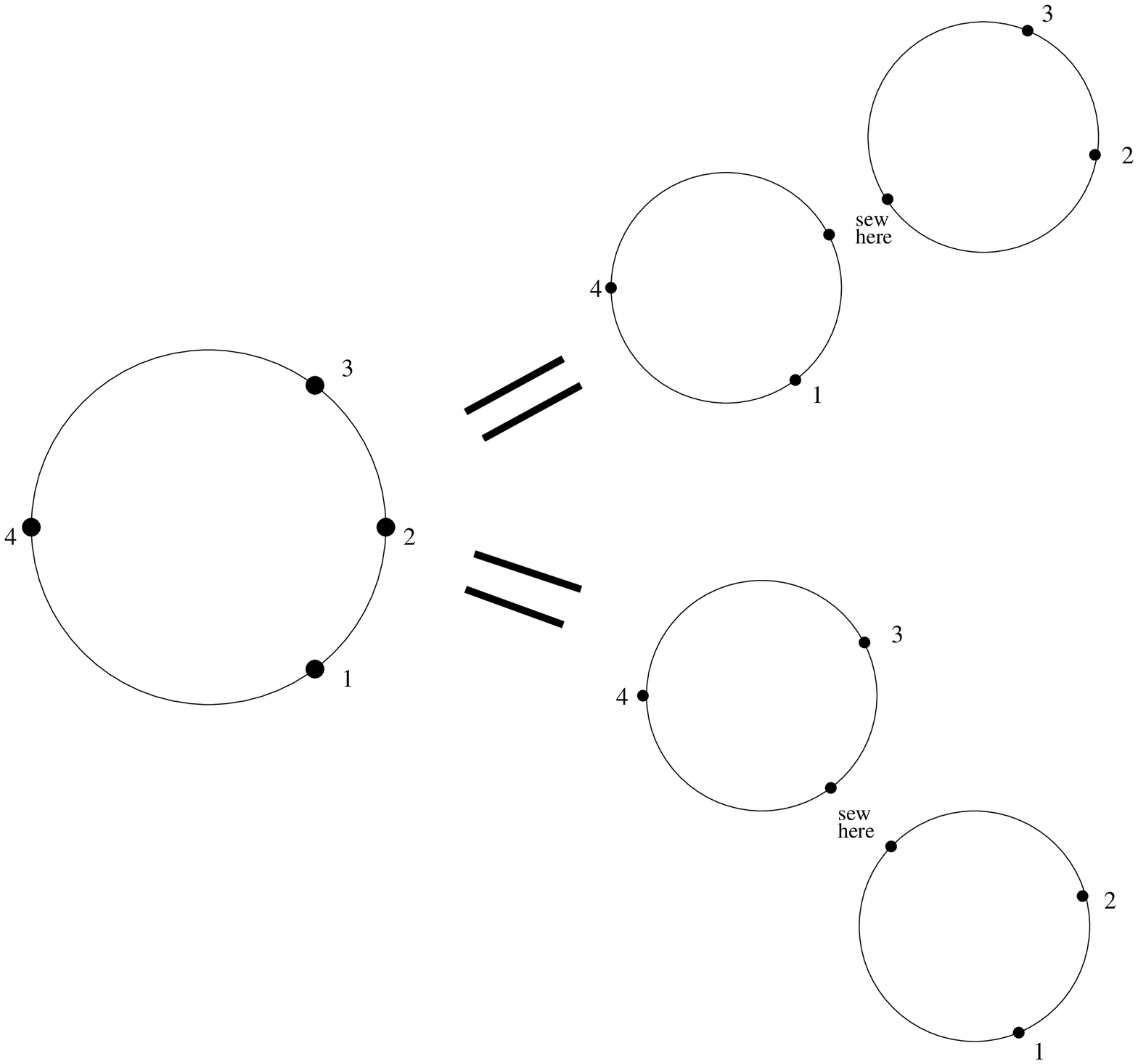,height=3in}}
\caption{The Relations of $\A$ Involving the Canonical Elements in \A(3)}
\label{associativity}
\end{figure}
\endgroup

\subsection{Deformations of associative algebras and Hochschild cohomology}

In this section, we shall review the manner in which an infinitesimal
deformation of an associative algebra gives rise to Hochschild cohomology.

Let $V$ be an $\A$-algebra with a morphism of operads $\mu:\A\, \to\,
\End{V}$.  Let $\rho := \mu_{\e{1}} = \id_V$, identity map on $V$,
and let the linear map $m := \mu_{\e{2}}:V^{\otimes 2}\,\to\,V$  let be
the {\sl multiplication operation}. Since $\A$ is a quadratic operad, $\mu$
is completely determined by compositions of $\rho$ and $m$. The associativity
relations (equation \ref{eq:assoc}) imply that $V$ is an associative algebra,
\ie\ the multiplication satisfies
\begin{equation}
m\oo_1 m = m\oo_2 m.\label{eq:assocalg}
\end{equation}
We shall now deform the associative algebra $V$ infinitesimally. Let
$\mu\,\mapsto\,\mu' = \mu + t \alpha  $ where $t$ is a small
parameter and $\alpha$ an element in $\End{V}(2)$ then impose that $\mu'$
make $V$ into an associative algebra up to first order in $t$.  Since $\mu'$
is entirely determined by $m':=\mu'_{\e{2}}$ (and the unit element which
remains fixed), we need only impose associativity of $m'$ (equation
\ref{eq:assocalg}, replacing $m$ by $m'$) keeping terms of first order in
$t$. Putting all terms on one side of the equality, we obtain
\begin{equation}
m\oo_2\alpha -\alpha\oo_1 m + \alpha\oo_2 m -m\oo_1 \alpha  = 0.
\label{eq:twococycle}
\end{equation}
However, it is possible that the infinitesimal deformation $\alpha$, up to
first order in $t$, gives rise to an algebra which is isomorphic to the
original, \ie\ that the deformation is trivial. This occurs if there exists a
$\beta$ in $\End{V}(1)$ such that the map $\Phi\,:=\,\id_V - t \beta $
satisfies
\begin{equation}
 m'\oo(\Phi\otimes\Phi) = \Phi\oo m.
\end{equation}
Plugging in $m' = m + t\alpha $ and $\Phi = {\id}_V - t\beta + $ and keeping
terms up to first order in $t$, we obtain
\begin{equation}
\alpha = m\oo_2\beta -\beta\oo_1 m + m\oo_1\beta \label{eq:twocoboundary}
\end{equation}
Therefore, (nontrivial) infinitesimal deformations of $V$ are classified by
$\alpha$ in $\End{V}(2)$ satisfying equation \ref{eq:twococycle} quotiented
by those which satisfy \ref{eq:twocoboundary} for some $\beta$ in $\End{V}(1).$
Equation $\ref{eq:twococycle}$ is precisely the condition that
$\alpha$ be a $2$-cocycle in Hochschild cohomology of $V$ with values in
itself while equation $\ref{eq:twocoboundary}$ is precisely the condition
that $\alpha$ be a $2$-coboundary in $V$ with values in itself.  Therefore,
the space of nontrivial infinitesimal deformations of $V$ is determined by
dimension two cohomology classes of $V$ with values in itself. Noting that
equations $\ref{eq:twococycle}$ and $\ref{eq:twocoboundary}$ make sense even
if $\alpha$ were an element of $\Hom(V^{\otimes 2},M)$ where $M$ is a
$V$-module, we are led to the following definition of the Hochschild
cohomology of the associative algebra $V$ with values in a $V$-module
$M$. The space of Hochschild  $n$-cochains is $\C{\A}{n}(V,M) :=
\Hom(V^{\otimes n},M)$ and the differential $d: \C{\A}{n}(V,M) \,\to\,
\C{\A}{n+1}(V,M)$, whose actions on $1$ and $2$ cochains are given by the
right hand sides of equations $\ref{eq:twococycle}$ and
$\ref{eq:twocoboundary}$, is defined by
\begin{equation}
d\alpha := m\oo_2\alpha +\sum_{i=1}^n \,(-1)^i\,\alpha\oo_i m + (-1)^{n+1}
m\oo_1 \alpha.\label{eq:hochdifferential}
\end{equation}
It is a simple exercise to verify that $d^2 = 0$ from the fact that $V$ is
an associative algebra and $M$ a $V$-module. Figure $\ref{differential}$
illustrates the geometric meaning of the terms in the Hochschild differential
of the $3$-cochain, $a$, namely.
$$ da = m\oo_1 a - a\oo_1 m + a\oo_2 m - a\oo_3 m + m\oo_2 a$$
We view $da$ as being associated to the canonical element in $\A(4)$ and the
terms in the differential denote various decompositions of the canonical
element in $\A(4)$ into the canonical element in $\A(3)$ and $\A(2)$.
The multiplication operation $m$ is associated with the canonical element
in $\A(2)$ and $a$ is associated with the canonical elements in $\A(3)$ which
are composed as indicated.
\begingroup
\input{psfig}
\begin{figure}[h]
\centerline{\psfig{figure=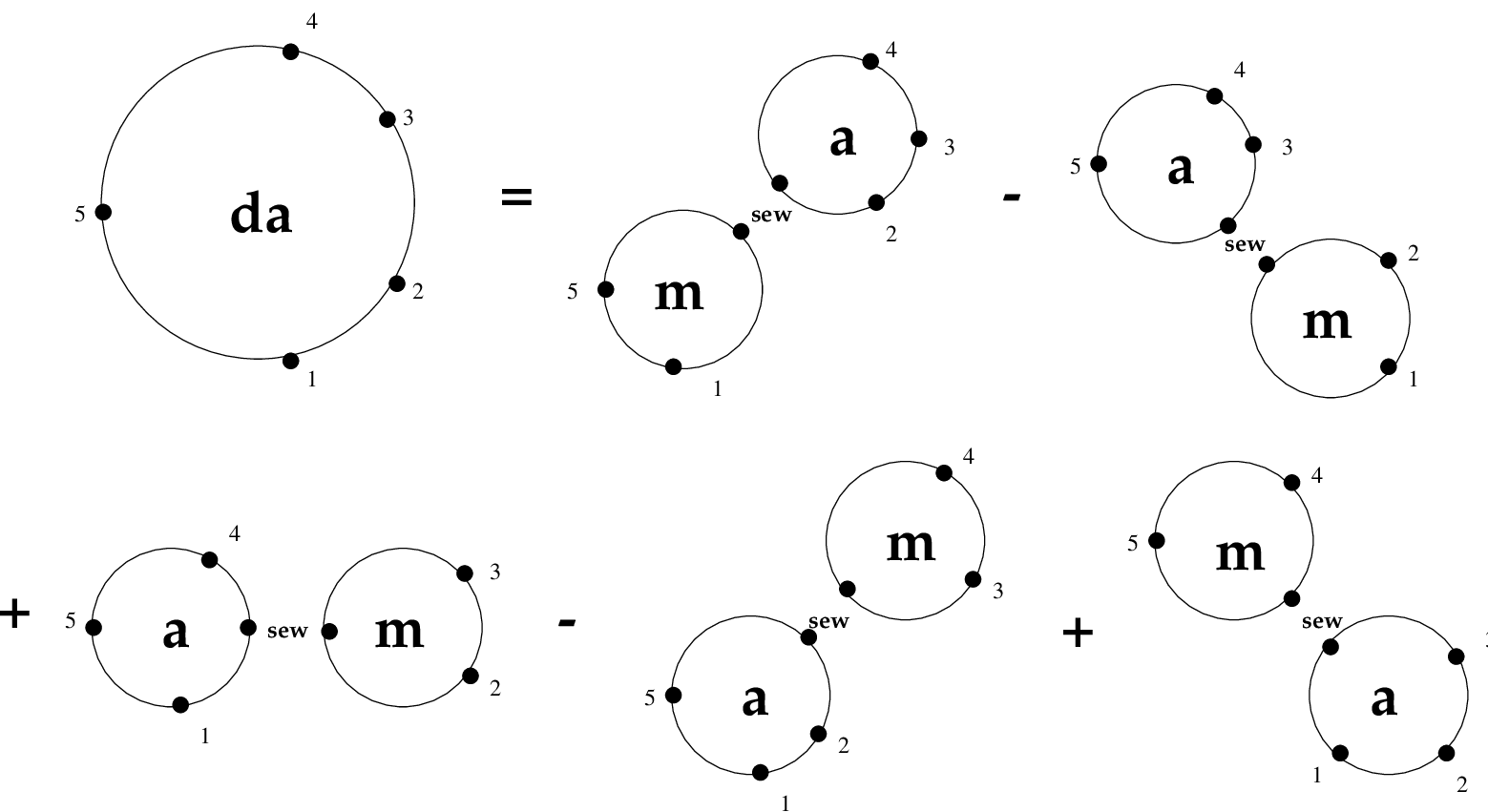,height=3in}}
\caption{The Differential of a $3$-cocycle}
\label{differential}
\end{figure}
\endgroup
We have just shown the following.
\begin{prop}
Let $V$ be an associative algebra (an $\A$-algebra) and $M$ a $V$-module then
the associated cochain complex, $(\C{\A}{\bullet}(V,M),d),$ is the Hochschild
complex of $V$ with values in $M$. Furthermore, the inequivalent,
infinitesimal deformations of $V$ are classified by elements in
$H_{\A}^2(V,V)$.
\end{prop}

\subsection{The little intervals operad}

A geometric operad which is well-known from homotopy theory (see \cite{m})
is the {\sl little intervals operad}, $\D = \{\,\D(n)\,\}_{n\geq 1}$. $\D(n)$
is the space of configurations of $n$ distinct, ordered embeddings of $I$
into $I$ where each embedding is a composition of a dilation (multiplication
by a positive real number) and a translation (addition of a real number),
where the embedded intervals are allowed to intersect at their
boundaries. $S_n$ acts on $\D(n)$ by reordering the $n$ maps and the
composition maps $\D(n)\times\D(n')\,\to\,\D(n+n'-1)$ which take $(\Sigma,
\Sigma') \,\mapsto \Sigma\oo_i\Sigma'$ are given by taking the large interval
of $\Sigma'$ and identifying the large interval with the $i$-th interval in
$\Sigma$ after shrinking it down to size via a dilation. It will prove useful
to construct another realization of this operad.

Let $\RP{1}$ inherit an orientation from $\nr^2$ so that it is diffeomorphic
to an oriented circle.  The homogeneous coordinates on $\RP{1}$ gives rise to
a standard coordinate, $x$, which identifies $\RP{1}$ with
$\nr\cup\{\,\infty\,\}.$  Consider the moduli space of distinct, ordered,
real analytic, orientation preserving embeddings $f_i:I\,\to\,\RP{1}$ for $i=
1, \ldots, n+1$ whose images are mutually disjoint except, possibly, at their
boundaries and identify any two such configurations if they are related by an
orientation preserving real projective transformation, \ie\ an element of
$\PSL$ which acts on $\RP{1}$ via
\begin{equation*}
x  \mapsto \, \frac{ax+b}{cx+d},\ \forall\, a,b,c,d\in\nr\  \text{satisfying}\
\det\left(\matrix a & b\\ c & d \endmatrix\right) = 1.
\end{equation*}
The resulting space, $\G(n)$, for all $n\geq 1$, forms an operad. The
permutation group acts by permuting the ordering on the first $n$ maps. The
composition maps $\G(n)\,\times\,\G(n')\,\to\,\G(n+n'-1)$ which
takes $(\Sigma, \Sigma')\,\mapsto\,\Sigma\oo_i\Sigma'$ are obtained by
cutting out the $(n'+1)$st interval of $\Sigma'$ and the $i$th interval of
$\Sigma$ and sewing them together along their boundaries in an orientation
preserving fashion. This operad is a real analytic analog of the operad
governing conformal field theory. This operad has a suboperad which is
isomorphic to the little intervals operad.

Consider an element $(f_1,\ldots,f_n)$ of $\D(n)$ and identify the large
interval in an orientation preserving way with the unit interval in the
standard coordinate about $x=0$ in $\RP{1}$ and let $f_{n+1}:I\, \to\,\RP{1}$
be an orientation preserving identification of $I$ with the unit interval in
the standard coordinate about $x=\infty$. Let $\B(n)$ consist of those points
in $\G(n)$ which arise from configurations of $(n+1)$ distinct, ordered
orientation preserving embeddings of $I$ into $\RP{1}$ of the type just
described. The operad $\B,$ which is isomorphic to the little intervals
operad, is a suboperad of $\G$. The operad $\A$ shares many features in
common with $\B$ (in fact, $\A$ may be regarded as a quotient of the space of
elements in $\B$ arising from nonintersecting intervals) but $\B$ has a much
richer geometric structure than $\A$. $\B(n)$ is a $2n$ dimensional manifold
with corners with $n!$ components where the  components are permutated into
one another by the free action of $S_n$, {\sl i.e.}\ $\B(n)\,\to\,\B(n)/S_n$
is a trivial $S_n$ bundle. However, there is a canonical section of this
bundle provided by the orientation of $\RP{1}$ in analogy to $\A$, \ie\
assign to each orbit of $S_n$ in $\B(n)$ the unique point consisting of maps
whose ordering increases as one proceeds around $\RP{1}$ as prescribed by its
orientation. Let $\Bh(n)$ denote the image of the canonical section in
$\B(n)$ which we shall identify with $\B(n)/S_n$. $\Bh$ is closed under the
composition maps in analogy with the associative case. For convenience, we
shall denote $\B(1) = \Bh(1)$ by $\K$ which is a Lie semigroup with unit
$\identity$.

Just as in the associative operad, the operad $\B$ is quadratic or, more
precisely, the linear operad generated by $\B$ is quadratic. All elements in
$\B(n)$ are obtained from elements in $\K$ and $\B(2)$ by compositions and
the action of the permutation group. However, unlike $\A$, $\K$ contains more
than one element -- it is a two dimensional manifold with corners. The
composition maps endow $\B(n)$ with the structure of a
$(\K,\K^n)$-space. $\B$-algebras are determined not only by elements in
$\B(2)$ but also by nontrivial elements in $\K$. To be more specific, let
$\Sigma$ be in a point in $\Bh(3)$ then, in analogy with equation
\ref{eq:assoc} and figure \ref{associativity} for the associative case, there
exist elements $\Sigma_1$, $\Sigmat_1$, $\Sigma_2$, and $\Sigmat_2$ in
$\Bh(2)$ such that
\begin{equation}
\Sigma = \Sigmat_1\oo_1\Sigma_1 = \Sigmat_2\oo_2 \Sigma_2
\label{eq:bassociativity}
\end{equation}
where $\Sigma_i$ is an element in $\Bh(2)$ which can be represented by a
circle containing intervals $i$ and $(i+1)$ from the original circle $\Sigma$
while $\Sigmat_i$ contains the remaining two intervals of the original
$\Sigma$. However, unlike the associative operad, the decomposition of
$\Sigma$ into $\Sigma_i$ and $\Sigmat_i$ is not unique because there are
nontrivial elements in $\K$. Choose $i=1,2$ and $\Sigma$ in
$\Bh(3)$ and let $W_i(\Sigma)$ be the set of all pairs $(\Sigmat_i,\Sigma_i)$
in $\Bh(2) \times\Bh(2)$ such that $\Sigma =
\Sigmat_i\oo_i\Sigma_i$. Define an equivalence relation on $W_i(\Sigma)$ by
demanding that the relations be generated by $(\Sigma,k\oo_1\Sigma') \sim
(\Sigma\oo_i k,\Sigma')$ for all $k$ in $\K$. In this case, $W_i(\Sigma)/\sim$
consists of one element which means that up to the appropriate action of
$\K$, there is a unique way to decompose a given element in $\Bh(3)$ into
a pair of elements in $\Bh(2)$ for a given $i$. The relations of the operad
$\B$ are generated by the free action of $S_3$ upon equation
\ref{eq:bassociativity}.

\subsection{Deformations of algebras over the little intervals operad}
We shall now obtain the cohomology associated to a $\B$-algebra by studying
the infinitesimal deformations of algebras over it.

We say that $V$ is a $\B$-algebra if  $V$ is a topological  vector space with
a smooth morphism of operads $\mu:\B\,\to\,\End{V}$. In particular, a
$\B$-algebra $V$ is a $\K$-module with binary operations parametrised by
$\B(2)$ which obey quadratic relations. The deformation problem that we will
consider is to characterize infinitesimal inequivalent deformations of a
given $\B$-algebra $V$ keeping the $\K$-module structure on $V$ fixed.

As in the associative case, the equivariance under the permutation group
means that $\mu$ is completely determined by its restriction to $\Bh$. Let
$\rho:\K\,\to\, \End{V}(1)$ (denoted by $k\,\mapsto\,\rho_k$) be the
restriction of $\mu$ to $\K$ and $m:\Bh(2)\,\to\,\End{V}(2)$ (denoted by
$\Sigma\,\mapsto\,m_\Sigma$) be the restriction of $\mu$ to $\Bh(2)$. $\rho$
makes $V$ into a $\K$-module. Because $\mu$ is quadratic, $\rho$ and $m$
completely determine $\mu$. The operations $m$ and $\rho$ must satisfy,
for all $k,k'$ in $\K$, and $\Sigma$ in $\Bh(3),$ $\Sigma_i$, $\Sigmat_i$,
$\Sigma'$ in $\Bh(2),$ equation \ref{eq:bassociativity} and
\begin{enumerate}
\item ($\K$-module structure of $V$) $\rho_{k\oo_1 k'} =
\rho_k\oo_1\rho_{k'},$
\item ($(\K,\K^2)$-equivariance of $\Bh(3)$)
\begin{enumerate}
\item $m_{k\oo_1\Sigma'} = \rho_k\oo_1 m_{\Sigma'}$
\item $m_{\Sigma'\oo_1 k} = m_{\Sigma'}\oo_1\rho_k,$
\item $m_{\Sigma'\oo_2 k} = m_{\Sigma'}\oo_2\rho_k,$ and
\end{enumerate}
\item (associativity) $m_{\Sigmat_1}\oo_1 m_{\Sigma_1} = m_{\Sigmat_2} \oo_2
m_{\Sigma_2}.$
\label{eq:relations}
\end{enumerate}
Consider a smooth map $\alpha :\Bh(2)\,\to\,\End{V}(2)$ which
infinitesimally deforms $m$ to $m' := m + t\alpha$ for small $t$.
Replacing $m$ by $m'$ in the associativity condition \ref{eq:relations} and
keeping terms first order in $t$ and putting all terms to one side of the
equation then one finds that $\alpha$ must be $(\K,\K^2)$-equivariant and
satisfy
\begin{equation}
m_{\Sigmat_2} \oo_2 \alpha_{\Sigma_2} - \alpha_{\Sigmat_1}\oo_1 m
_{\Sigma_1} + \alpha_{\Sigmat_2} \oo_2 m_{\Sigma_2} - m_{\Sigmat_1}\oo_1
\alpha _{\Sigma_1} = 0
\label{eq:btwococycle}
\end{equation}
for all $\Sigma$ in $\Bh(3),$ $\Sigma_i$, $\Sigmat_i$ in $\Bh(2)$ satisfying
equation \ref{eq:bassociativity}. This is the $2$-cocycle condition which
is analogous to equation $\ref{eq:twococycle}$ in Hochschild cohomology.
Suppose that the infinitesimal deformation $\alpha$, of first order in
$t$, gives rise to a trivial deformation. This occurs if there exists a map
$\beta$ in $\End{V}(1)$ commuting with the $\K$-module structure such that
$\Phi\,:=\,\id_V - t \beta$ satisfying
\begin{equation}
m'_{\Sigma}\oo(\Phi\otimes\Phi) = \Phi \oo m_{\Sigma}
\label{eq:bautomorphism}
\end{equation}
for all $\Sigma$ in $\Bh(2).$ Plugging in $m' = m + t\alpha$ and $\Phi =
{\id}_V - t\beta$ and keeping terms up to first order in $t$, we obtain
\begin{equation}
\alpha_\Sigma = m_\Sigma\oo_2\beta - \beta\oo_1 m_\Sigma + m\oo_1\beta
\label{eq:btwocoboundary}.
\end{equation}
Equation \ref{eq:btwococycle} is the condition that $\alpha$ is a two
cocycle and equation \ref{eq:btwocoboundary} is the condition that $\alpha$
is a two coboundary in the cohomology of a $\B$-algebra $V$ with values in
itself. Let $M$ be a $V$-module then for $n\geq 1$, the space of $n$-cochains
of $\B$-cohomology of $V$ with values in a $V$-module $M$ $\C{\B}{n}(V,M)$, is
defined to be the set of all smooth maps $\alpha:\Bh(n)\,\to\,\Hom(V^{\otimes
n},M)$  which are $(\K,\K^n)$ equivariant. We also define $\C{\B}{0}(V,M)
:= M.$ The differential $d: \C{\B}{n}(V,M)\, \to\,\C{\B}{n+1}(V,M)$, as
suggested by equations $\ref{eq:btwococycle}$ and $\ref{eq:btwocoboundary}$,
is defined by the formula, for $n\geq 2$,
\begin{equation}
(d\alpha)_{\Sigma} = m_{\Sigma_0}\oo_2\alpha_{\Sigmat_0} + \sum_{i=1}^n\,
(-1)^n\, \alpha_{\Sigmat_i}\oo_i m_{\Sigma_i} + (-1)^{n+1}
m_{\Sigma_{n+1}}\oo_1 \alpha_{\Sigmat_{n+1}}\label{eq:bdifferential}
\end{equation}
where $\Sigma$ is an element of $\Bh(n+1)$, $\Sigma_i$ are elements of
$\Bh(2)$, $\Sigma_i$ are elements of $\Bh(n-1)$, such that for all
$i=1,\ldots, n$,
\begin{equation}
\Sigma = \Sigmat_i\oo_i\Sigma_i = \Sigma_0\oo_2 \Sigmat_0 = \Sigma_{n+1}\oo_1
\Sigmat_{n+1}.
\end{equation}
The expression for the differential is well-defined because of the
$(\K,\K^n)$-equivariance of $\alpha$. The differential of an element $\beta$
in $\C{\B}{1}(V,M)$ is given by
\begin{equation}
(d\beta)_\Sigma = m_\Sigma\oo_2\beta_\identity - \beta_\identity\oo_1 m_\Sigma
+ m_\Sigma\oo_1 \beta_\identity
\end{equation}
for all $\Sigma$ in $\B(2)$. Finally, the differential of an element $\gamma$
in $\C{\B}{0}(V,M)$ is given by
\begin{equation}
(d\gamma)_\Sigma = m_\Sigma\oo_2\gamma - m_\Sigma\oo_1\gamma
\end{equation}
for all $\Sigma$ in $\K$.

Notice that if $\beta'$ belongs to $\C{\B}{1}(V,M)$ then
$(\K,\K)$-equivariance means that
\begin{equation}
\beta'_k = \rho_k\oo_1\beta'_e = \beta'_e\oo_1\rho_k
\end{equation}
for all $k$ in $\K$ so $\beta'$ is completely determined by $\beta'_e$ which
is an element of $\Hom(V,M)$. It is this $\beta := \beta'_\identity$ which
appears in equation $\ref{eq:btwocoboundary}$.

It is straightforward to verify that $d^2 = 0$ from the associativity
condition $\ref{eq:bassociativity}$. Figure $\ref{differential}$, once again,
graphically demonstrates the meaning of the various terms in the differential
of a $3$-cochain, $a$, where the $\Sigma_i$'s correspond to the circles about
$m$ and $\Sigmat_i$ correspond to the circles about $a$. We have just shown
the following.

\begin{sloppypar}
\begin{th}
Let $\B$ be the little intervals operad, $V$ a $\B$-algebra, and $M$ a
$V$-module. The associated cochain complex is $(C_{\B}(V,M),d).$ Furthermore,
inequivalent infinitesimal deformations of $V$ as a $\B$-algebra are
classified by its second cohomology, $H_{\B}^2(V,V)$.
\end{th}
\end{sloppypar}

In the special case where $V$ is a $\B$-algebra with a trivial $\K$-module
structure then notice that the $(\K,\K^2)$ equivariance of $m$
implies that $V$ is nothing more than an associative algebra and the complex
$\C{\B}{\bullet}(V,M)$ can be identified with the Hochschild complex of $V$
with values in $M$.

\subsection{A complex of vertex operator algebras}
The little intervals operad, $\B$, embeds into the operad which governs
CFT's. Therefore, any CFT is a $\B$-algebra. In this section, we shall write
down, explicitly, the complex described in the previous section in the
context of vertex operator algebras. This gives rise to a complex written
purely in terms of vertex operator algebras which closely resembles
Hochschild cohomology.

There is a morphism of operads $\B\, \hookrightarrow\, \p$ defined as
follows. The canonical inclusion $\nr^2\,\hookrightarrow\,\nc^2$ induces an
inclusion  $\RP{1}\,\hookrightarrow\,\CP{1}$. Given any element $\Sigma$ in
$\B(n)$ represented by embedded intervals, $(f_1,\ldots, f_{n+1}),$ in
$\RP{1}$, then each $f_i$ can be naturally regarded as an embedding of $I$
into $\CP{1}$ which, by analytic continuation, becomes a biholomorphic
embedding of $D$ into $\CP{1}$. The resulting Riemann sphere with
holomorphically embedded disks is an element in $\p(n).$ Therefore, if $V$ is
a CFT then it is necessarily a $\B$-algebra and its infinitesimal
deformations as a $\B$-algebra are governed by the complex defined in the
previous section.

Elements in $\B(2)$ are specified by the inclusion maps $(f_1,f_2)$ where
$f_1(x) = a_1 x + z_1$ and $f_2(x) = a_2 x + z_2$ for some positive numbers
$a_1$ and $a_2$ (dilations) as well as real numbers $z_1$ and $z_2$
(translations). We shall denote such an element of $\Bh(2)$ by $(z_1, a_1,
z_2, a_2)$. If $V$ arises from a vertex operator algebra then the
multiplication operation $m:\Bh(2)\, \to\,\End{V}(2)$ is given by,
for all $v_1$ and $v_2$ in $V$,
\begin{equation}
\label{eq:explicit}
m_{(z_1,a_1,z_2,a_2)}(v_1\otimes v_2) := \cases Y((a_1)^{-L_0} v_1,z_1)
Y((a_2)^{-L_0} v_2,z_2)\vacuum,&\text{if $|z_1| > |z_2|$}\\
Y((a_2)^{-L_0} v_2,z_2) Y((a_1)^{-L_0} v_1,z_1) \vacuum,&\text{if $|z_2| >
|z_1|$}\endcases
\end{equation}
and everywhere else $m$ is defined by analytic continuation. $L_0$ is the
usual element in the Virasoro algebra and $\vacuum$ is the vacuum vector, the
image of the element in $\p(0)$ with the standard chart about $\infty$. This
can be used to explicitly write down the differential in $\C{\B}{\bullet}.$
An element $\alpha$ in $\C{\B}{n}(V,M)$ can now be written as
$\alpha_{(z_1,a_1,\ldots,z_n a_n)}$ which must be $(\K, \K^n)$ equivariant,
using similar coordinates to the above.

Enlarge the spaces $\G(n)$ to a space $\GG(n)$ which is the moduli space of
$(n+1)$ distinct, orientation preserving, real analytic coordinates in $\RP{1}$
such that no two coordinates have centers which coincide and where any two such
are identified if they are related by an action of $\PSL$. The collection
$\GG := \{\,\GG(n)\,\}_{n\geq 1}$ forms a partial operad where the
composition maps are given by the one dimensional analogue of sewing in
$\pp$. Clearly, $\G$ is a suboperad of the partial operad $\GG.$

There is an important partial suboperad of $\GG$, $\cc$, where $\cc(n)$ is
diffeomorphic to the configuration space of $n$ points on $\nr$. That is, let
$\cc(n)$ consist of $(n+1)$ distinct, ordered orientation preserving
embeddings, $(f_1,\ldots, f_{n+1}),$ of $I$ into $\RP{1}$ where for all
$i=1,\ldots, n$, $f_i$ is embedded into $\nr$ by translations only and
$f_{n+1}$ is the standard unit interval about $\infty$ so that no two have
centers which coincide. $\cc := \{\,\cc(n)\,\}_{n\geq 1}$ forms a partial
operad which is quadratic in the sense that every element in $\cc(n)$ can be
obtained from compositions of elements in $\cc(2)$ and the action of the
permutation group. In fact, the cohomology governing deformations of a
$\cc$-algebra naturally give rise to the a complex for the cohomology of a
$\cc$-algebra, $V$, with values in a the $V$-module, $M$,
$d:\C{\cc}{n}(V,M)\,\to\,\C{\cc}{n+1}(V,M)$  where $n$-cochains consist of
smooth, $(\cc(1)^n,\cc(1))$-equivariant maps from $\cch(n)\,\to\,\Hom(V^n,M)$
where $\cch(n)$ is the canonical component of $\cc(n)$. If $\alpha$ is an
element in $\C{\cc}{n}$ and $(z_1,\ldots,z_n)$ is a point in $\cch(n)$ then
the differential is given by
\begin{multline}
(d\alpha)_{(z_1,\ldots,z_{n+1})} = m_{(z_1,0)}\oo_2
\alpha_{(z_2, \ldots, z_{n+1})} \\ +  \sum_{i=1}^n\,(-1)^i\,
\alpha_{(z_1,\ldots,z_i,z_{i+2}, \ldots z_{n+1})}\oo_i m_{(0,z_{i+1} - z_i)}
+ (-1)^{n+1} m_{(0,z_{n+1})} \oo_1 \alpha_{(z_1,\ldots,z_n)}
\end{multline}
where $m$ denotes the binary multiplication on $V$ or the module action of
$V$ on $M$ as it appropriate. We have shown the following.
\begin{th}
Let $V$ be a $\cc$-algebra and $M$, a $V$-module then
$(\C{\cc}{\bullet}(V,M),d)$ is the associated cochain complex which governs
deformations of $\cc$-algebras.
\end{th}

In the case where $V$ is a vertex operator
algebra and $v_1$, $v_2$ belong to $V$, $m$ is given by
\begin{equation}
m_{(z_1,z_2)}(v_1\otimes v_2) := \cases Y(v_1,z_1)
Y(v_2,z_2)\vacuum,&\text{if $|z_1| > |z_2|$}\\ Y(v_2,z_2) Y(v_1,z_1)
\vacuum,&\text{if $|z_2| > |z_1|$}\endcases
\end{equation}
and is defined by analytic continuation elsewhere.

In the definition of a vertex operator algebra, the most important axiom is
the so-called associativity axiom. Formally deforming the map $Y$
infinitesimally, we are led to an expression that is precisely that which is
written above.

\begin{ack}
We would like to thank the organizers of the Texel Conference for their
hospitality. We are also grateful to the Summer Geometry Institute in Park
City, Utah for their hospitality and support where part of this work was
done. We would like to thank Giovanni Felder, Jose Figueroa-O'Farrill, Dan
Freed, Yi-Zhi Huang, Maxim Kontsevich, Yuri I. Manin, Andrey Reznikov, Albert
Schwarz, Yan Soibelman, Jim Stasheff,  Arkady Vaintrob and Edward Witten for
useful conversations, both electronic and aural. Our special thanks to Pierre
Deligne for pointing out some flaws in our usage of dual spaces in an earlier
version of the manuscript.
\end{ack}

\bibliographystyle{amsplain}

\makeatletter \renewcommand{\@biblabel}[1]{\hfill#1.}\makeatother
\ifx\undefined\bysame
\newcommand{\bysame}{\leavevmode\hbox to3em{\hrulefill}\,}
\fi

\end{document}